\documentclass[a4paper]{article}[12pt]
\usepackage{amsfonts}
\usepackage{amsmath,cite,tikz,multirow,amssymb,mathrsfs,latexsym,bm,inputenc,longtable}
\usepackage{tabularx}
\usepackage[figuresright]{rotating}

\begin{document}

\newcommand{\qed}{\hphantom{.}\hfill $\Box$\medbreak}
\newcommand{\Proof}{\noindent{\bf Proof \ }}

\newtheorem{theorem}{Theorem}[section]
\newtheorem{lemma}[theorem]{Lemma}
\newtheorem{corollary}[theorem]{Corollary}
\newtheorem{remark}[theorem]{Remark}
\newtheorem{example}[theorem]{Example}
\newtheorem{definition}[theorem]{Definition}
\newtheorem{construction}[theorem]{Construction}
\newtheorem{proposition}[theorem]{Proposition}

\title{\large{\bf The existence of non-classical orthogonal quantum Latin squares\footnote{Supported by Natural Science Foundation of Hebei Province under Grant A2023205045 (Zihong Tian) and Science Foundation of Hebei Normal University  under Grant L2024B52 (Yajuan  Zang).}}}

\author{Yan Han$^{a}$, Yajuan Zang$^{b}$, Hongjiao Zhang$^{a}$, Zihong Tian$^{ac}$\thanks{Corresponding author. E-mail address: tianzh68@163.com}\\
\small $^a$ School of Mathematical Sciences, Hebei Normal University,
           Shijiazhuang 050024,  China\\
\small $^b$ Department of Elementary Education, Hebei Normal University,
          Shijiazhuang 050024,  China\\
\small $^c$  Hebei Research Center of the Basic Discipline Pure Mathematics,
  Shijiazhuang 050024,   China}

\date{}

\maketitle

\noindent {\bf Abstract:} Quantum Latin squares are a generalization of classical Latin squares in
quantum field and have wide applications in unitary error bases, mutually unbiased
bases, $k$-uniform states and quantum error correcting codes. In this paper, we put
forward   some new  quantum Latin squares with special properties, such as  idempotent quantum Latin square, self-orthogonal quantum Latin square,  holey quantum Latin square, and the notions of orthogonality on them. We present some  forceful construction methods including PBD constructions and   filling in holes constructions  for non-classical quantum Latin squares. As  consequences, we establish   the existence of non-classical 2-idempotent MOQLS$(v)$, non-classical 2, 3-MOQLS$(v)$ and non-classical SOQLS$(v)$ except possibly for several definite values.\vspace{0.1cm}


\section{Introduction}

Latin square is an important combinatorial configuration, which is  an very active research topic, and has great applications in many practical fields such as  statistics, cryptography, computer science, communications  and bioinformatics \cite{Donald,Hayashi,Gao,Pal,Laywine,Ryazanov}.

Let $[v]=\{0,1,\ldots, v-1\}$. A \emph{Latin square} of order $v$, LS$(v)$, is a  $v \times v$ array $L=(L_{i, j})$ in which each of the elements  of $[v]$ occurs exactly once in each row and  column. A Latin square $L$ of order $v$ is \emph{idempotent} if $L_{i, i} = i$ for $ i\in [v]$. Two Latin squares $L_1,L_2$ of order $v$ are \emph{orthogonal}, if when $L_1$ is superimposed on $L_2$, every ordered pair $00, 01, \ldots,v - 1 v - 1$ occurs. Especially, a \emph{self-orthogonal} Latin square (SOLS) is a Latin square that is orthogonal to its transpose. A set of Latin squares $L_1, L_2,\ldots,L_t$ is \emph{mutually orthogonal}, or a set of MOLS, if for every $1 \leq i < j\leq t$, $L_i$ and $L_j$ are orthogonal, denoted by $t$-MOLS$(v)$. Especially, a set of idempotent Latin squares $L_1, L_2,\ldots,L_t$ is \emph{mutually orthogonal}, denoted by $t$-idempotent MOLS$(v)$.  We use $N(v)$ to represent the maximum number  of Latin squares in a set of MOLS of  order $v$, and use $N(1^v)$ to represent the maximum number of Latin squares in a set of idempotent MOLS of order $v$. Latin squares have been extensively studied  and have a wealth of results.

\begin{lemma}\label{md}{\rm(\!\!\cite{Bose,Bush,Colbourn,Hedayat,Todorov})}\label{lem1.1}
$(1)$ If $q$ is a prime power, then $N(q)= q-1;$

$(2)$ $v = p_{1}^{r_1}p_{2}^{r_2}\ldots p_{s}^{r_s} $, where $s\geq 2$, $p_i$ is a prime and $p_i\neq p_j$ for $1\leq i\neq j\leq s$, $r_i$ is a positive integer, then $N(v)\geq min\{p_{i}^{r_i}-1 : 1 \leq i \leq s\};$

$(3)$ For any positive integer $v\neq 2,6$, $N(v)\geq 2;$

$(4)$ For any positive integer $v\neq 2,3,6,10$, $N(v)\geq 3;$

$(5)$ For any positive integer $v\neq 2,3,4,6,10,22$, $N(v)\geq 4;$

$(6)$  For any positive integer $v\neq 2,3,6,$ there exists an SOLS$(v);$

$(7)$ For any positive integer $v\neq 2,3,6$, $N(1^v)\geq 2;$

$(8)$ $N(v)-1\leq N(1^v) \leq N(v)$.

\end{lemma}

 Quantum Latin square is   a quantum analog of Latin square, which is introduced by Vicary et al. in 2016 \cite{Musto1}.
Quantum Latin squares have connections to other primitives used in quantum information, such as unitary
error bases \cite{Musto1}, mutually unbiased bases \cite{Musto2}, $k$-uniform states \cite{Goyeneche0},  controlled families of Hadamards \cite{Reutter},   quantum teleportation and quantum error correction \cite{Cheng,Musto2,Preskill}, etc.

 A \emph{quantum Latin square}  $\Phi$ of dimension $v$, denoted by QLS$(v)$, is a $v \times v$  array of elements $|\Phi_{i,j}\rangle$ of the complex vector space $\mathbb{C}^{v}$, $i,j\in [v]$, such that every row and every column determine an orthonormal basis of $\mathbb{C}^{v}$.
 Two quantum Latin squares $\Phi$, $\Psi$ of dimension $v$ are \emph{equivalent} if there exists some unitary operator $U$ on $\mathbb{C}^{v}$, family of modulus 1 complex numbers $c_{ij}$  and permutations $\sigma, \tau\in S_{v}$, such that   $\left| \Psi_{ij} \right\rangle $ $= c_{ij}U\left| \Phi_{\sigma(i),\tau(j)} \right\rangle $ for all $i,j \in [v]$,  where  $S_{v}$ is the  symmetry group of degree $v$.

 We call the orthonormal basis $\{|0\rangle,|1\rangle,\ldots,|v-1\rangle\}$ of $\mathbb{C}^{v}$ a \emph{computational basis}. Actually, if place the vector $|l\rangle$ instead of $l$ in a Latin square of order $v$, we get a quantum Latin square for which the elements in every row and column form a computational basis, which is called a \emph{classical} quantum Latin square. The equivalent of a classical quantum Latin square is also called a classical quantum Latin square, otherwise it is called a \emph{non-classical} quantum Latin square. Obviously, if we change the computational basis into an any orthonormal basis of  $\mathbb{C}^{v}$ in a classical quantum Latin square, then the new quantum Latin square is also classical.
  Unless otherwise specified, we index the rows and columns of a QLS($v$) by computational basis of $\mathbb{C}^{v}$.

Two quantum Latin squares $\Phi$, $\Psi$ of dimension $v$ are \emph{orthogonal}, just when the set of vectors
$\{|\Phi_{ij}\rangle\otimes |\Psi_{ij}\rangle: i,j\in [v ] \} $ forms an orthonormal basis of the space $\mathbb{C}^{v}\otimes \mathbb{C}^{v}$,
i.e., $((|\Phi_{ij}\rangle\otimes |\Psi_{ij}\rangle),(|\Phi_{i'j'}\rangle\otimes |\Psi_{i'j'}\rangle))=
\langle\Phi_{ij}|\Phi_{i'j'}\rangle\langle\Psi_{i,j}|\Psi_{i'j'}\rangle=\delta_{ii'}\delta_{jj'}$ for $i,j,i',j'\in [v]$.
A set of $t$ quantum Latin squares of dimension $v$, say $\Phi_1, \Phi_2,\ldots, \Phi_t$, are said to be \emph{mutually
orthogonal} if $\Phi_i$ and $\Phi_j$  are orthogonal for all $1 \leq i < j \leq t$, denoted by $t$-MOQLS($v$).
 Especially, a \emph{self-orthogonal} quantum Latin square (SOQLS) is a quantum Latin square that is orthogonal to its conjugate transpose.

Since the introduction of QLS by Vicary et al., an QLS has been extensively studied and has yielded some significant results. So far, there are the following results about non-classical MOQLSs.

\begin{lemma}\label{Md1}{\rm (\!\!\cite{Zang0,Paczos,Squares,Boyadzhiyska})}
$(1)$ Suppose that $v = p_{1}^{r_1}p_{2}^{r_2}\ldots p_{s}^{r_s} $, where $s\geq 2$, $p_i$ is a prime and $p_i\neq p_j$ for $1\leq i\neq j\leq s$, $r_i$ is a positive integer, then there exists a non-classical $t$-MOQLS$($$v$$)$, where $t= min_{1\leq i\leq s}\{p^{r_i}_i-1\}$$;$
moreover, if $s=1,r_1\geq 2$, with $r_1=r'+r''$, $0< r'\leq r''$, then there exists a non-classical $t$-MOQLS$($$v$$)$, where~ $t=p^{r'}_1-1$.

$(2)$ Let $E_1=\{2,3,4,6,8,18,p,2p,6p \}$, if $v\not \in E_1 $, where $p\geq 5$ is an odd prime, then there exists a non-classical $2$-MOQLS$($$v$$)$.

$(3)$ Let $E_2=\{9,12,24,27,50,54,3p\}$, if $v\not \in E_1\cup E_2$, where $p\geq 5$ is an odd prime, then there exists a non-classical $3$-MOQLS$($$v$$)$.

$(4)$ Let $E_3=\{16,32,36,48,66,110,242,4p\}$, if $v\not \in E_1\cup E_2\cup E_3$, where $p\geq 5$ is an odd prime, then there exists a non-classical $4$-MOQLS$($$v$$)$.

$(5)$ If $d_1,d_2\geq 4$, there exists a non-classical SOQLS$($$d_1d_2$$)$, where $d_1d_2  \neq36$.
\end{lemma}

Similar to the classical idempotent Latin square, we can give the corresponding definition of idempotent quantum Latin square. A quantum Latin square $\Phi$ of  dimension $v$ is \emph{idempotent} if the diagonal elements $|{{\Phi}_{ii}} \rangle$   for $i \in [v]$  form an orthonormal basis  of  ${{\mathbb C}^{v}}$. Further, a set of idempotent quantum Latin squares $\Phi_1, \Phi_2,\ldots, \Phi_t$ is mutually orthogonal, denoted by $t$-\emph{idempotent} MOQLS$(v)$.

In this paper, we mainly study the constructions and the existence of non-classical $t$-MOQLS, $t$-idempotent MOQLS and SOQLS by applying combinatorial design theory.

The rest of the paper organized as follows. In Section 2, we present some constructions of  non-classical quantum Latin squares by using  some auxiliary designs including pairwise balanced designs  and   incomplete quantum Latin squares.   In Section 3, we establish   the existence of non-classical 2-idempotent MOQLS$(v)$, non-classical 2, 3-MOQLS$(v)$ and non-classical SOQLS$(v)$
except possibly for several definite values. Finally, we draw our conclusions and outlook.

\section{Constructions}
In this section, we propose two construction methods for non-classical quantum Latin squares  based on pairwise balanced designs and  incomplete quantum Latin squares. Before start, we introduce some statements. Unless otherwise specified, we index the rows and columns by the computational basis of $\mathbb{C}^v$ in a QLS($v$). For vector spaces $U$, $V$ and $W$, we use the symbol ``$U \oplus V$'' to represent the direct sum of spaces $U$ and $V$, and use the symbol ``$W \ominus V$'' to represent the vector space $U$ with ``$U \oplus V= W$''.
\subsection{PBD constructions}

Let $K$~be a set of positive integers, $X$~a finite set of size~$v$ and $\mathcal B$~a family of subsets (called $blocks)$ of~$X$. If~$B\in \mathcal B$,~$|B|\in K$, and every pair of different elements in $X$ occurs in exactly one block of $\mathcal B$, then ($X$,~$\mathcal B$)~is called a \emph{pairwise balanced design}  (PBD), denoted by PBD$(v, K)$.

PBD is a kind of classical combinatorial configuration, which has rich results and wide applications in combinatorial design theory.

\begin{lemma}{\rm (\!\!\cite{Colbourn})}\label{PBD1}
$(1)$ When~$K=\{3,4,5\}$,  $v\notin\{6,8\}$, there exists a PBD$(v, K)$.

$(2)$ When $K=\{4,5,7,9,10,11\}$,  $v\notin\{6,8,12,14,15,18,19,23,26,27,30\}$, there exists a PBD$(v, K)$.
\end{lemma}

\begin{construction}\label{PBD2}
If there exists a PBD$(v, K)$  and there exists an idempotent QLS$(k)$ for any  $k\in K$,  then there exists a non-classical  idempotent QLS$($$v$$)$.
\end{construction}

\Proof Suppose $([v], \mathcal{B})$  is a PBD$(v, K)$. Take the   element $0$ and let $\mathcal{B}_0=\{B: 0\in B, B \in \mathcal{B}\} $, $\mathcal{B}'=\mathcal{B}\backslash\mathcal{B}_0$.

For   $B=\{0, x_1, \ldots, x_{k-1}\}\in\mathcal{B}_0$, we construct an idempotent QLS$(k)$ indexed rows and columns by $|0\rangle, |{{x}_{1}}\rangle, \ldots, |{{x}_{k-1}}\rangle$ on subspace ${\mathcal{L}}_{B}=span\{|0\rangle, |{{x}_{1}}\rangle, \ldots,  |{{x}_{k-1}}\rangle\}$ of $\mathbb{C}^{v}$, denoted by ${{M}_{B}}=\{|{M}_{B}(i,j)\rangle: i, j\in\{0, x_1,  \ldots, x_{k-1}\}\}$, where  $|{M}_{B}(i,j)\rangle $ is the vector $|0\rangle$ or is from a non-computational basis of the subspace $span\{|{{x}_{1}}\rangle,  \ldots, |{{x}_{k-1}}\rangle\}$ of $\mathbb{C}^{v}$.

For   $B=\{y_1, y_2, \ldots, y_{l}\}\in\mathcal{B}'$, we construct a classical idempotent QLS$(l)$ indexed rows and columns by $|{{y}_{1}}\rangle, |{{y}_{2}}\rangle, \ldots, |{{y}_{l}}\rangle$ on subspace
 ${\mathcal{L}}_{B}=span\{|{y_1}\rangle, |{y_2}\rangle, $ $ \ldots, |{y_{l}}\rangle\}$ of $\mathbb{C}^{v}$,
denoted by ${{N}_{B}}=\{|{N}_{B}(i,j)\rangle :i, j\in\{y_1, y_2, \ldots, y_{l}\}\}$, where  $|{N}_{B}(i,j)\rangle$ is from the set $\{|{y_1}\rangle,   |{y_2}\rangle, \ldots, |{y_{l}}\rangle\}$.

Define a $v\times v$ array ${{\Phi }}=({|{{\Phi }}(x,y)\rangle})$, where
$$ |{{\Phi }}(x,y)\rangle =\left \{
\begin{array}{ll}
|M_{B}(x,y)\rangle,   & x\neq y, x,y\in B, B\in \mathcal{B}_0;\\
 |N_{B}(x,y)\rangle,   & x\neq y, x,y\in B, B\in \mathcal{B}';\\
 |M_{B}(x,x)\rangle,   & x\in B, B \in \mathcal{B}_0.
\end{array} \right. $$
Then  ${{\Phi }}$ is a non-classical idempotent QLS($v$).

First, ${{\Phi }}$ is a   QLS($v$). For any row $|{{x}}\rangle$, columns $|{y}_{1}\rangle, |{y}_{2}\rangle$, $y_1\neq y_2$, we   need to prove that
\begin{equation}\label{formula 1.1}
\langle{{\Phi }}(x,{y}_{1})|{{\Phi }}(x,{y}_{2})\rangle=0.
\end{equation}

By the definition of PBD, there exist  blocks ${B}_1, {B}_2 \in\mathcal{B}$  such that $\{x, {y}_{1}\}\subseteq {{B}_{1}}$, $\{x, {y}_{2}\}\subseteq {{B}_{2}}$. When $B_1=B_2$, then $\{x, y_1, y_2\}\subseteq B_1$. So $|{{\Phi }}(x,{y}_{1})\rangle, |{{\Phi }}(x,{y}_{2})\rangle$   come  from the same idempotent QLS  $M_{B_1}$ or $N_{B_1}$, we have  $\langle{{\Phi }}(x,{y}_{1})|{{\Phi }}(x,{y}_{2})\rangle=0.$
When ${{B}_{1}}\ne{{B}_{2}}$, ${{B}_{1}\cap{B}_{2}}$ =$\{x\}$, then ${\mathcal{L}}_{{B}_{1}}\cap{\mathcal{L}}_{{B}_{2}}=$ span$\{|{x}\rangle\}$. In this case, we can know that if\vspace{0.2cm}

$(1)$ ${{B}_{1}}, {{B}_{2}}\in\mathcal{B}'$, we have $(|{{\Phi }}(x,{y}_{1})\rangle , |{{\Phi }}(x,{y}_{2})\rangle )=0$, that is to say  Eq.~(\ref{formula 1.1}) holds.

$(2)$ ${{B}_{1}}\in\mathcal{B}_0, {{B}_{2}}\in\mathcal{B}'$, we have $|{{\Phi }}(x,{y}_{2})\rangle\in {\mathcal{L}}_{{B}_{2}}\ominus$ span$\{|{x}\rangle\}$, so  Eq.~(\ref{formula 1.1}) holds.

$(3)$ ${{B}_{1}}, {{B}_{2}}\in\mathcal{B}_0$, then $x=0$, and $|0\rangle$   appears in row $|0\rangle$ and column $|0\rangle$, so  Eq.~(\ref{formula 1.1}) holds.\vspace{0.2cm}

Second, ${{\Phi }}$ is  idempotent. For any $x_1, x_2\in[v]$, $x_1\neq x_2$,  we need to prove that
\begin{equation}\label{formula 1.2}
\langle{{\Phi }}(x_1,{x}_{1})|{{\Phi }}(x_2,{x}_{2})\rangle=0.
\end{equation}

In fact, if $\{x_1, x_2\}\subseteq B, B\in\mathcal{B}$, Eq.~(\ref{formula 1.2}) holds as $|{{\Phi }}(x_1,{x}_{1})\rangle, |{{\Phi }}(x_2,{x}_{2})\rangle$  come  from the same idempotent QLS $M_B$ or $N_B$.

If $x_1\in B_1, x_2\in B_2$, $B_1\neq B_2$, since $\bigcup_{B\in \mathcal{B}_0}(B \backslash\{0\})=[v]\backslash \{0\}$ by the definition of PBD, we assume  $B_1, B_2\in\mathcal{B}_0$ and $B_1\cap B_2 =\{0\}$. Further,  we know  the subspace $\mathcal{L}_{B_1}\cap\mathcal{L}_{B_2}=$ span$\{|0\rangle \}$ and $\bigoplus_{B\in\mathcal{B}_0}(\mathcal{L}_B\ominus $ span$\{|0\rangle \})=\mathbb{C}^{v}\ominus$ span$\{|0\rangle \}$. From idempotent QLSs $M_{B_1}$ and $M_{B_2}$,  $|{{\Phi }}(x_1,{x}_{1})\rangle, |{{\Phi }}(x_2,{x}_{2})\rangle$  are orthogonal. So Eq. (\ref{formula 1.2}) is true.

In addition, $\Phi$ is non-classical, since the elements are from computational basis or non-computational basis, which ends the proof.
\qed

\begin{construction}\label{con2.4}
If there exists a PBD$(v, K)$  and there exists a $t$-idempotent  MOQLS$(k)$ for any $k\in K$,   then there exists a non-classical $t$-idempotent MOQLS$(v)$.
\end{construction}
\Proof  Suppose $([v], \mathcal{B})$  is a PBD$(v, K)$. Take the element $0$ and let $\mathcal{B}_0=\{B: 0\in B, B \in \mathcal{B}\} $, $\mathcal{B}'=\mathcal{B}\backslash\mathcal{B}_0$.

For any~$B=\{0, x_1, \ldots, x_{k-1}\}\in\mathcal{B}_0$, we construct a $t$-idempotent MOQLS$(k)$ indexed rows and columns by $|0\rangle, |{{x}_{1}}\rangle, \ldots, |{{x}_{k-1}}\rangle$ on the subspace ${\mathcal{L}}_{B}=span\{|0\rangle, |{{x}_{1}}\rangle, $ $\ldots,  |{{x}_{k-1}}\rangle\}$ of $\mathbb{C}^{v}$, denoted by ${{M}_{B}^{f}}=\{|{M}_{B}^{f}(i,j)\rangle: i, j\in\{0, x_1, \ldots, x_{k-1}\}\}, 1\leq f\leq t$, where  $|{M}_{B}^{f}(i,j)\rangle$ is $|0\rangle$ or from a non-computational basis of the subspace $span\{|{{x}_{1}}\rangle, |{{x}_{2}}\rangle, \ldots, |{{x}_{k-1}}\rangle\}$ of $\mathbb{C}^{v}$.

For any $B=\{y_1, y_2, \ldots, y_{l}\}\in\mathcal{B}'$, we construct a $t$-idempotent MOQLS$(l)$ indexed rows and columns by~$|{{y}_{1}}\rangle, |{{y}_{2}}\rangle, \ldots, |{{y}_{l}}\rangle$ on the subspace
 ${\mathcal{L}}_{B}=span\{|{y_1}\rangle,  |{y_2}\rangle, \ldots, |{y_{l}}\rangle\}$ of $\mathbb{C}^{v}$,
denoted by ${{N}_{B}^{f}}=\{|{N}_{B}^{f}(i,j)\rangle : i, j\in\{y_1, y_2, \ldots, y_{l}\}\}, 1\leq f\leq t$, where  $|{N}_{B}^{f}(i,j)\rangle $ is from a computational basis of the subspace ${\mathcal{L}}_{B}$ of $\mathbb{C}^{v}$.

For $1\leq f\leq t$, define a $v\times v$ array  ${{\Phi }^{f}}=({|{{\Phi }^{f}}(x,y)\rangle})$, where
$$ |{{\Phi }}^{f}(x,y)\rangle =\left \{
\begin{array}{ll}
|M_{B}^{f}(x,y)\rangle,   & x\neq y, x,y\in B, B\in \mathcal{B}_0;\\
 |N_{B}^{f}(x,y)\rangle,   & x\neq y, x,y\in B, B\in \mathcal{B}';\\
 |M_{B}^{f}(x,x)\rangle,   & x\in B, B \in \mathcal{B}_0.
\end{array} \right.$$

Similar to the proof of Construction \ref{PBD2}, we know ${{\Phi }^{f}}$ is a non-classical idempotent QLS($v$). Next, we prove ${{\Phi }^{f}}$,  $1\leq f\leq t$, are   mutually orthogonal.

For any ${{f}_{i}}\ne {{f}_{j}}, 1\leq i\neq j \leq t, ({x}_{1},{y}_{1})\ne ({x}_{2}, {y}_{2})$, we only need to prove that

\begin{equation}
\langle{{\Phi }^{{{f}_{i}}}}({{x}_{1}},{{y}_{1}})|{{\Phi }^{{{f}_{i}}}}({{x}_{2}},{{y}_{2}})\rangle~\langle{{\Phi }^{{{f}_{j}}}}({{x}_{1}},{{y}_{1}})|{{\Phi }^{{{f}_{j}}}}({{x}_{2}},{{y}_{2}})\rangle=0. \label{Eq.5}
\end{equation}

When $\{{{x}_{1}},{{y}_{1}}\}, \{{{x}_{2}},{{y}_{2}}\}\subseteq B$, $B \in \mathcal{B}$, since $|{M}_{B}^{f}(i,j)\rangle $  or $|{N}_{B}^{f}(i,j)\rangle$, $1\leq f\leq t$, is a  $t$-idempotent MOQLS($|B|$), so Eq. (\ref{Eq.5})  holds.

When $\{{{x}_{1}},{{y}_{1}}\}\subseteq {{B}_{1}}, \{{{x}_{2}},{{y}_{2}}\}\subseteq {{B}_{2}}$, $B_1\neq B_2,$ $B_1, B_2 \in \mathcal{B}$, if
 $\langle{{\Phi }^{{{f}_{i}}}}({{x}_{1}},{{y}_{1}})|{{\Phi }^{{{f}_{i}}}}({{x}_{2}},{{y}_{2}})\rangle$ $\neq 0$, then  there exists an element $x_m \in [v]$ such that ${\mathcal{L}}_{{B}_{1}}\cap{\mathcal{L}}_{{B}_{2}} = span\{|{{x}_{m}}\rangle \}$; if $\langle{{\Phi }^{{{f}_{j}}}}({{x}_{1}},{{y}_{1}})|{{\Phi }^{{{f}_{j}}}}({{x}_{2}},\\{{y}_{2}})\rangle \ne 0$, then there exists an element $x_n \in [v]$ such that ${\mathcal{L}}_{{B}_{1}}\cap{\mathcal{L}}_{{B}_{2}}=span\{|{{x}_{n}}\rangle \}$, and $x_m\neq x_n$. So, $span\{|{{x}_{m}}\rangle,|{{x}_{n}}\rangle\}\subseteq {\mathcal{L}}_{{B}_{1}}\cap{\mathcal{L}}_{{B}_{2}}$, there is a contradiction
as  ${{B}_{1}}, {{B}_{2}}$ are  the blocks of  PBD$(v, K)$ and $|{{B}_{1}}\cap{{B}_{2}}|\leq1$. Therefore, Eq. (\ref{Eq.5}) holds.
\qed
\subsection{Filling in holes constructions}

$~\vspace{-0.4cm}$

We first introduce some relevant auxiliary designs,   then  we give a non-classical MOQLS by using the filling in holes constructions.

Suppose $S$ is a finite set of size $v$, and   $H = \{S_1,S_2,\ldots,S_n\}$ is a set of disjoint subsets of $S$, where $|S_i|=s_i$, $1\leq i \leq n$. A \emph{holey  Latin  square}  HLS$(v; s_1, s_2,\ldots , s_n)$ of order $v$ with hole set $H$ is a  $v\times v$ array $L$ whose rows and columns are indexed by the elements of $S$, and satisfies the following properties: \vspace{0.1cm}

$(1)$ each cell of $L$ is either empty or contains an element of $S$;

$(2)$ the subarrays indexed by $S_i \times S_i$ are empty (these subarrays are \emph{holes});

$(3)$ the elements in row or column $s$ are exactly those of $S\setminus S_i$ if $s \notin S_i$, and of $S$ otherwise.\vspace{0.1cm}

\noindent
 We will say that the type of $L$ is the multiset $\{|S_1|,|S_2|,\ldots,|S_n|\}$. We also use the notation $s_1^{n_1}s_2^{n_2} \ldots s_l^{n_l}$ to describe the type of an HLS, where  $\sum {_{i=1}^{l}}n_i=n$.

If $H=\{S_1\}$, then a holey Latin square is called an \emph{incomplete Latin square}, denoted by ILS($v; s_1$). If $H = \{S_1, S_2, \ldots, S_n\}$ with type $s_1^{n_1}s_2^{n_2} \ldots s_l^{n_l}$ is a partition of $S$, then a holey Latin square is called a \emph{partitioned holey Latin square}, denoted by PHLS($s_1^{n_1}s_2^{n_2} \ldots s_l^{n_l}$).

Two holey Latin squares on symbol set $S$ and hole set $H$, say $L_1$ and $L_2$, are said to be $orthogonal$ if their superimposition yields every ordered pair  in $(S \times S)\setminus \bigcup_{i=1}^{n}(S_i \times S_i)$. Similarly, $t$-HMOLS$(v; s_1, s_2,\ldots , s_n)$ denotes a set of $t$  HLS$(v; s_1, s_2,\ldots , s_n)$s that are  pairwise orthogonal \cite{Colbourn}. If $t$ HLSs of order $v$ being pairwise orthogonal are all ILS($v;s_1$)s or all PHLS($s_1^{n_1}s_2^{n_2} \ldots s_l^{n_l}$)s, then we denote $t$-IMOLS($v; s_1$) or $t$-PHMOLS($s_1^{n_1}s_2^{n_2} \ldots s_l^{n_l}$) for short respectively.

A holey Latin square is called $self$-$orthogonal$, denoted by HSOLS($v; s_1, s_2,\ldots , s_n$), if it is orthogonal to its transpose. If $H=\{S_1\}$ with $|S_1|=s_1$, or $H$ is a partition of $S$, then a holey SOLS is an ISOLS($v; s_1$) or a PHSOLS($s_1^{n_1}s_2^{n_2} \ldots s_l^{n_l}$).

It is clear that a PHLS$(1^v)$ is equivalent to an idempotent LS$(v)$ and a PHMOLS$(1^v)$ is equivalent to an idempotent MOLS$(v)$.

Here we list some results of HMOLSs.
\begin{lemma}\label{IOLS}{\rm(\!\!\cite{Bennett0,Colbourn,Denes,Heinrich})}
$(1)$ For  any positive integer $v\geq 4,$ there exists an ILS$(v;2)$.

$(2)$ For any positive integer $v>k \geq 1,$  there exists a $2$-IMOLS$(v;k)$ if and only if $v\geq 3k $ and $(v,k)\neq (6,1)$.

$(3)$ For any positive integer $v>k \geq 1,$ there exists a $3$-IMOLS$(v;k)$ if and only if $v\geq 4k,$   $(v,k)\neq (6,1),$ and possibly when $(v,k)= (10,1)$.

$(4)$ For any positive integer $v \geq 13,$ there exists an ISOLS$(v;4)$.
\end{lemma}

 Let $V= \{V_1, V_2, \ldots, V_n\}$ be a set of  mutually orthogonal subspaces of   $\mathbb{C}^v$. A $holey~quantum~\\ Latin ~square$ HQLS$(v; v_1, v_2,\ldots , v_n)$ of dimension $v$ with $hole~ set$ $V$ is a  $v \times v$ array $\Psi$ whose rows and columns are indexed by an orthogonal basis of $\mathbb{C}^v$, $\{\phi_1,\phi_2,\ldots,\phi_v\}$, where dim$V_i=v_i$ for $1\leq i \leq n$, and satisfies the following properties:\vspace{0.1cm}

$(1)$ every cell of $\Psi$ is either empty or contains a unit vector of $\mathbb{C}^v$;

$(2)$ the subarrays whose rows and columns indexed by the basis of $V_i$s are empty (these subarrays are \emph{holes});

$(3)$ suppose the row or column is indexed by $\phi$, then the elements in the row or column are exactly the basis of  $\mathbb{C}^v\ominus V_i$ if $\phi \in V_i$, and of
$\mathbb{C}^v $ otherwise.\vspace{0.1cm}

Similar with classical ones, we will say that the type of an HQLS is the multiset $\{$dim$V_1$,dim$V_2$,\\$\ldots$,dim$V_n\}$.
We also use the notation $v_1^{n_1}v_2^{n_2} \ldots v_l^{n_l}$ to describe the type of an HQLS, where  $\sum {_{i=1}^{l}}n_{i}=n$.

If $V=\{V_1\}$, then a holey quantum Latin square is called an \emph{incomplete quantum Latin square}, denoted by IQLS($v;v_1$). If $V= \{V_1, V_2, \ldots, V_n\}$ with $\oplus_{i=1}^n V_i=\mathbb{C}^v$, then a holey quantum Latin square is called a \emph{partitioned holey quantum Latin square}, denoted by PHQLS($v_1^{n_1}v_2^{n_2} \ldots v_l^{n_l}$).

A   holey quantum Latin square for which every element of the array is in the form of computational basis,  then  it  is called a \emph{holey classical quantum Latin square}.

Two holey quantum Latin squares on $\mathbb{C}^v$  and hole set $V$, say $\Psi$ and $\Phi$, are said to be $orthogonal$ if their ``superimposition" yields an orthonormal basis of $(\mathbb{C}^v \otimes \mathbb{C}^v)\ominus (\oplus_{i=1}^{n}(V_i \otimes V_i))$. $t$-HMOQLS$(v; v_1, v_2,\ldots , v_n)$ denotes a set of $t$ HQLS$(v; v_1, v_2,\ldots , v_n)$s that are  pairwise orthogonal. If $t$ IQLSs of dimension $v$ being pairwise orthogonal are all incomplete quantum Latin squares with a hole of dimension $v_1$ or all PIQLSs, then we denote $t$-IMOQLS($v;v_1$) or $t$-PHMOQLS($v_1^{n_1}v_2^{n_2} \ldots v_l^{n_l}$) for short respectively.

A holey quantum Latin square is called $self$-$orthogonal$ if it is orthogonal to its conjugate transpose. We use the notation HSOQLS$(v; v_1, v_2,\ldots, v_n)$ to denote   holey self-orthogonal quantum Latin square. If $V=\{V_1\}$ with dim$V_1=v_1$, or $V$ is a direct sum decomposition of $\mathbb{C}^v$, then a holey SOQLS is an ISOQLS($v;v_1$) or HSOQLS($v_1^{n_1}v_2^{n_2} \ldots v_l^{n_l}$).

Similarly, we replace the quantum Latin squares in the above definition with idempotent quantum Latin squares, then we can give their corresponding idempotent definitions.

In \cite{Zang0}, the authors listed filling in holes constructions just based on HSOQLS. Now, we will give more general filling in holes constructions based on IMOQLS and ISOQLS.

\begin{construction}\label{IQOLS}{\rm(\!\!\cite{Zang0})}
If there exists an HLS$($$v; v_1,v_2,\ldots,v_n$$)$ and an LS$(v_s)$  for $1\leq s \leq n$, then there exists a non-classical QLS$($$v$$)$.
\end{construction}

\begin{construction}\label{OLS1}
If there exists a $2$-HMOLS$(v;v_1,v_2,\ldots,v_n)$ and a $2$-MOLS$(v_s)$ for $1\leq s \leq n$, then there exists a non-classical $2$-IMOQLS($v;v_n$) and a non-classical $2$-MOQLS$(v)$.
\end{construction}
\Proof Suppose $V= \{V_1, V_2, \ldots, V_n\}$ is a set of  mutually orthogonal subspaces of  $\mathbb{C}^v$. Let $\phi^1$, $\phi^2$ be a classical 2-HMOQLS$(v;v_1,v_2,\ldots,v_n)$ with hole set $V$ based on computational basis of $\mathbb{C}^v $. Suppose $\psi_s^1$, $\psi_s^2$ is a pair of classical 2-MOQLS$(v_s)$ based on non-computational basis of the space $V_s$, $1\leq s\leq n-1$. Fill each $\psi_s^1$ and $\psi_s^2$ in $\phi^1$, $\phi^2$ respectively as a hole for $1\leq s\leq n-1$, then we can get two new squares  $\varphi^{1}=(|\varphi_{ij}^{1}\rangle)$ and $\varphi^{2}=(|\varphi_{ij}^{2}\rangle)$. It is not difficult to check that  $\varphi^{1}$ and $\varphi^{2}$ are both IQLS($v;v_n$)s. Here we just prove the orthogonality.

It is necessary to show that $\{|\varphi_{ij}^{1}\rangle\otimes |\varphi_{ij}^{2}\rangle: i,j \in  [v] \}$  yields an orthonormal basis of $(\mathbb{C}^v \otimes \mathbb{C}^v )\ominus  (V_n\otimes V_n)$, i.e.,
for any $i,j,k,l\in [v]$, we just need to prove the following equation
\begin{equation}
\langle\varphi^1_{ij}|\varphi^1_{kl}\rangle\langle\varphi^2_{ij}|\varphi^2_{kl}\rangle=0. \label{aaa}
\end{equation}
We   consider the following three cases.

Case 1. Suppose $(i,j)$,$(k,l)$ are the positions both not in a hole or both in the same hole. According to the orthogonality of $\phi^1$ and $\phi^2$, or $\psi_s^1$ and $\psi_s^2$, $1\leq s \leq n-1$,   Eq.~(\ref{aaa}) is established.

Case 2. Suppose $(i,j)$,$(k,l)$ are the positions in which the one is not in a hole and the other is in a hole. Without loss of generality, assume $(k,l)$ is the position in a hole, and $|\varphi^1_{kl}\rangle$, $|\varphi^2_{kl}\rangle$ are both in the subspace $V_s$  for some $1\leq s \leq n-1$. Suppose $\langle\varphi^1_{ij}|\varphi^1_{kl}\rangle\neq 0$ and $\langle\varphi^2_{ij}|\varphi^2_{kl}\rangle\neq 0$, then we have $|\varphi^1_{ij}\rangle$ and $|\varphi^2_{ij}\rangle$ are both in the subspace $V_s$. So we have $|\varphi^1_{ij}\rangle\otimes|\varphi^2_{ij}\rangle \in V_s\otimes V_s$ with contradiction. Thus   Eq. (\ref{aaa}) is established.

Case 3. Suppose $(i,j)$,$(k,l)$ are the positions respective from two different holes. Combining with the definition of holey quantum Latin square, we have $\langle\varphi^1_{ij}|\varphi^1_{kl}\rangle=\langle\varphi^2_{ij}|\varphi^2_{kl}\rangle=0$, then  Eq. (\ref{aaa}) is established.

Besides, $\varphi^1$ and $\varphi^2$ are non-classical,  since they are  filled with $\psi_s^1$ and $\psi_s^2$  both based on non-computational basis of space $V_s$.

As can be seen from above, we can get a 2-IMOQLS($v;v_n$) denoted by $\varphi^{1}$, $\varphi^{2}$.

Finally,  we fill  $\psi_{n}^1$ and $\psi_n^2$ in $\varphi^{1}$ and $\varphi^{2}$. Thus we obtain  two new quantum Latin squares as $\Phi^1$ and $\Phi^2$ by Construction \ref{IQOLS}. We assert that $\{\Phi^1, \Phi^2\}$ is a non-classical 2-MOQLS$(v)$.

In fact, for any $i,j,k,l\in [v]$, we just need to prove the following equation
\begin{equation}
\langle\Phi^1_{ij}|\Phi^1_{kl}\rangle\langle\Phi^2_{ij}|\Phi^2_{kl}\rangle=0. \label{bbb}
\end{equation}
We  consider the following cases.

Case 4. Suppose $(i,j)$,$(k,l)$ are the positions both  not in the hole or both in the hole of $\varphi^{1}$ and $\varphi^{2}$. According to the orthogonality of $\varphi^1$ and $\varphi^2$, or $\psi_n^1$ and $\psi_n^2$, we have   Eq.~(\ref{bbb}) is established.

Case 5. Suppose $(i,j)$,$(k,l)$ are the positions in which the one is not in the hole and the other is in the hole of $\varphi^{1}$ and $\varphi^{2}$. Without loss of generality, assume $(k,l)$ is the position from the hole, and $|\Phi^1_{kl}\rangle$, $|\Phi^2_{kl}\rangle$ are both in the subspace $V_n$. Suppose $\langle\Phi^1_{ij}|\Phi^1_{kl}\rangle\neq 0$ and $\langle\Phi^2_{ij}|\Phi^2_{kl}\rangle\neq 0$, then   $|\Phi^1_{ij}\rangle$ and $|\Phi^2_{ij}\rangle$ are both in the subspace $V_n$. So we have $|\Phi^1_{ij}\rangle\otimes|\Phi^2_{ij}\rangle \in V_n\otimes V_n$ with contradiction. Therefore,  Eq. (\ref{bbb}) is established.
\qed

This construction can be generalized to the case of $t$-HMOLSs. Moreover, if we use a non-classical 2-HMOQLS and 2-MOQLS instead of a classical 2-HMOLS and 2-MOLS in Construction \ref{OLS1}, the conclusions    also establish. It is just needed to change   Case 2 and   Case 5 in the proof of Construction \ref{OLS1} into   ``If $\langle\Phi^1_{ij}|\Phi^1_{kl}\rangle\neq 0$ and $\langle\Phi^2_{ij}|\Phi^2_{kl}\rangle\neq 0$, then   $span\{|\Phi^1_{ij}\rangle\} \cap V_s\neq \{0\}$ and $span\{|\Phi^2_{ij}\rangle\} \cap V_s\neq \{0\}$  for some $1\leq s \leq n$. So we have $span\{|\Phi^1_{ij}\rangle\otimes|\Phi^2_{ij}\rangle\} \cap (V_s\otimes V_s) \neq \{0\}$ with contradiction."

\begin{corollary}\label{IMOLS}
$(1)$ If there exists a $t$-HMOLS$(v;v_1,v_2,\ldots,v_n)$ and a $t$-MOLS$(v_s)$ for $1\leq s \leq n$, then there exists a non-classical $t$-MOQLS$(v)$.

$(2)$ If there exists a $t$-HMOQLS$(v;v_1,v_2,\ldots,v_n)$ and a $t$-MOQLS$(v_s)$ for $1\leq s \leq n$, then there exists a non-classical $t$-MOQLS$(v)$.
\end{corollary}

\begin{construction}\label{ISOLS1}
If there exists an HSOLS$(v;v_1,v_2,\ldots,v_n)$ and an SOLS$(v_s)$ for $1\leq s \leq n$, then there exists a non-classical SOQLS$(v)$.
\end{construction}
\Proof For $1\leq s \leq n-1$, according to Construction \ref{OLS1}, it is just needed to prove the self-orthogonality.
Suppose $V= \{V_1, V_2, \ldots, V_n\}$  is a set of  mutually orthogonal subspaces of $\mathbb{C}^v$. Let $\phi$ be a classical HSOQLS$(v;v_1,v_2,\ldots,v_n)$ with hole set $V$ based on computational basis of $\mathbb{C}^v$. Suppose $\psi_s$ is a classical SOQLS$(v_s)$ based on non-computational basis in the subspace $V_s$, $1\leq s\leq n-1$. Fill each $\psi_s$ in $\phi$ respectively as a hole for $1\leq s\leq n-1$ and we denote the new quantum Latin squares as $\Phi$. We assert that $\Phi$ is  a non-classical SOQLS$(v;v_n)$.

In fact, for any $i,j,k,l\in [v]$, we just need to prove the following equation
\hspace{2cm}\begin{equation}
\langle\Phi_{ij}|\Phi_{kl}\rangle\langle\Phi_{ji}|\Phi_{lk}\rangle=0. \label{12}
\end{equation}

We   consider the following three cases.

Case 1. Suppose $(i,j)$,$(k,l)$ are the positions both not in a hole or both in the same hole. According to the self-orthogonality of $\phi$ or $\psi_s$,   Eq.~(\ref{12}) is established.

Case 2. Suppose $(i,j)$,$(k,l)$ are the positions in which one is not in a hole and the other is in a hole. Without loss of generality, assume $(k,l)$ is the position in a hole, and $|\Phi_{kl}\rangle$, $|\Phi_{lk}\rangle$ are both in the subspace $V_s$ for some $1\leq s \leq n$. Suppose $\langle\Phi_{ij}|\Phi_{kl}\rangle\neq 0$ and $\langle\Phi_{ji}|\Phi_{lk}\rangle\neq 0$, then   $|\Phi_{ij}\rangle$ and $|\Phi_{ji}\rangle$ are both in the subspace $V_s$. So we have $|\Phi_{ij}\rangle\otimes|\Phi_{ji}\rangle \in V_s\otimes V_s$ with contradiction. Thus  Eq.~(\ref{12}) is established.

Case 3. Suppose $(i,j)$,$(k,l)$ are the positions respective from two different holes. Combining with the definition of incomplete quantum Latin square,~we have~$\langle\Phi_{ij}| \Phi_{kl}\rangle=\langle\Phi_{ji}|\Phi_{lk}\rangle=0$, so   Eq. (\ref{12}) is established.

As can be seen from above, $\Phi$ is a non-classical ISOQLS$(v;v_n)$.

Finally,  we can obtain a non-classical SOQLS$(v)$ by   filling  $\psi_{n}$  in $\Phi$.
\qed

If we replace  classical HSOLS and SOLS with   non-classical HSOQLS and SOQLS, the conclusion  in Construction \ref{ISOLS1} also holds.

\begin{corollary}
\hspace{0.1cm}If there exists an HSOQLS$(v;v_1,v_2,\ldots,v_n)$ and an SOQLS$(v_s)$ for $1\leq s \leq n$, then there exists a  non-classical SOQLS$(v)$.
\end{corollary}

Next, we use the   conjugate idempotent ILS to study   2-idempotent MOQLS. First, we introduce some related concepts.

Let $(Q,\odot)$ be a quasigroup. We  define six binary operations $\odot(1,2,3)$, $\odot(1,3,2)$, $\odot(2,1,3)$, $\odot(2,3,1)$, $\odot(3,1,2)$ and $\odot (3,2,1) $ on the set $Q$ as follows:\vspace{0.2cm}

$a\odot b=c$ if and only if
$a\odot(1,2,3)b=c$,~~~$a\odot(1,3,2)c=b$,~~~$b\odot(2,1,3) a=c$,

\hspace{3.5cm} $b\odot(2,3,1)c=a$,~~~$c\odot(3,1,2)a=b$,~~~$c\odot(3,2,1)b=a$.\vspace{0.2cm}

\noindent These six quasigroups~$(Q,\odot(i,j,k))$ are called \emph{conjugates} of the quasigroup $(Q,\odot)$.

It is well known that  the multiplication table of a quasigroup $(Q,\odot)$  defines a Latin square $L$, then   the six Latin squares defined by the  multiplication tables of its conjugates $(Q,\odot(i,j,k))$  are called \emph{conjugates} of $L$.

A Latin square which is orthogonal to its $(i,j,k)$-conjugate is called $(i,j,k)$-conjugate orthogonal Latin square, where $\{i,j,k\}=\{1,2,3\}$.

Similarly, an idempotent  ILS$(v;n)$ which is orthogonal to its $(i,j,k)$-conjugate is called an \emph{idempotent $(i,j,k)$-conjugate orthogonal incomplete Latin square},  denoted by idempotent $(i,j,k)$-COILS$(v;n)$.

\begin{lemma}\label{lem2.6}{\rm (\!\!\!\cite{Bennett})}
$(1)$ When $n\geq 1$, if $v\geq 8n+42$, there exists an idempotent $(3,2,1)$-COILS$(v;n)$.

$(2)$ When $n=4,5$,  $v\geq 3n+1$ and
 $(i)$ $n=4$, $v\notin\{14,15,18,19,23,27\};$
 $(ii)$ $n=5$,~$v\notin\{18,19,22,23,28,30,34\}$,  there exists an idempotent~$(3,2,1)$-COILS$($$v;n$$)$.
\end{lemma}

\begin{construction}\label{con2.7}
If there exist  a pair of  idempotent $(3,2,1)$-COILS$($$v;n$$)$ and  a $2$-idempotent MOQLS$(n)$, then there exists a non-classical $2$-idempotent MOQLS$(v)$.
\end{construction}

\Proof Let ${{L}_{1}}$ and ${{L}_{2}}$ be a pair of   idempotent $(3,2,1)$-COILS$(v;n)$ on $[v]$ with the hole $H=\{v-n,v-n+1,\dots,v-1\}$  at the lower right corner of  them,  and  $\phi=(|\phi(x,y)\rangle)$, $\psi=(|\psi(x,y)\rangle)$ be their corresponding classical quantum idempotent $(3,2,1)$-COILS$(v;n)$ composed of the computational basis of ${{\mathbb C}^{v}}$.

 Let ${{\phi}_{1}}=(|{{\phi}_{1}}(x,y)\rangle)$ and ${{\psi}_{1}}=(|{{\psi}_{1}}(x,y)\rangle)$ be a pair of classical orthogonal idempotent QLS($n$) based on the non-computational basis of the subspace $W=span\{|v-n\rangle, |v-n+1\rangle, \dots, |v-1\rangle\}$ of ${{\mathbb C}^{v}}$.

Define two $v\times v$ arrays $\Phi=(|\Phi(x,y)\rangle)$ and $\Psi=(|\Psi(x,y)\rangle)$  as follows:

$$ |\Phi(x,y)\rangle =\left \{
\begin{array}{ll}
|\phi(x,y)\rangle,   & x\in [v-n]~{\mbox{or}}~y\in [v-n];\\
 |{{\phi}_{1}}(x,y)\rangle,   & x,y\in [v]\backslash [v-n].
\end{array} \right.$$

$$ |\Psi(x,y)\rangle =\left \{
\begin{array}{ll}
|\psi(x,y)\rangle,   & x\in [v-n]~{\mbox{or}}~y\in [v-n];\\
|{{\psi}_{1}}(x,y)\rangle,   &x,y\in [v]\backslash [v-n].
\end{array} \right.$$
Then $\Phi$ and $\Psi$ both are non-classical idempotent QLS($v$)s by Construction \ref{IQOLS}. Here, we only need to prove the orthogonality i.e. the following equation holds for any $({{x}_{1}},{{y}_{1}})\ne ({{x}_{2}},{{y}_{2}})$

\begin{equation}\label{2.13}
\langle\Phi({{x}_{1}},{{y}_{1}})|\Phi({{x}_{2}},{{y}_{2}})\rangle \langle\Psi({{x}_{1}},{{y}_{1}})|\Psi({{x}_{2}},{{y}_{2}})\rangle=0.
\end{equation}

When ${{x}_{1}}, {{x}_{2}}\in[v-n]$ or ${{y}_{1}}, {{y}_{2}}\in[v-n]$, according to the orthogonality of $\phi$ and $\psi$,  Eq. (\ref{2.13}) holds. 
When ${{x}_{1}}, {{y}_{1}}\in[v]\backslash [v-n]$ and ${{x}_{2}}, {{y}_{2}}\in[v]\backslash [v-n]$, according to the orthogonality of $\phi_1$ and $\psi_1$, Eq. (\ref{2.13}) holds. 
When ${{x}_{1}}, {{y}_{1}}\in[v-n], {{x}_{2}}, {{y}_{2}}\in[v]\backslash [v-n]$, $\langle\Phi({{x}_{1}},{{y}_{1}})|\Phi({{x}_{2}},{{y}_{2}})\rangle\langle\Psi({{x}_{1}},{{y}_{1}})|\Psi({{x}_{2}},$ ${{y}_{2}})\rangle
=\langle\phi({{x}_{1}},{{y}_{1}})|{{\phi}_{1}}({{x}_{2}},{{y}_{2}})\rangle\langle\psi({{x}_{1}},{{y}_{1}})|{{\psi}_{1}}({{x}_{2}},$ ${{y}_{2}})\rangle$. If $\langle\phi({{x}_{1}},{{y}_{1}})|{{\phi}_{1}}({{x}_{2}},{{y}_{2}})\rangle\ne0$, according to the orthogonality of $\phi$ and $\psi$,   then $|\psi({{x}_{1}},{{y}_{1}})\rangle\in {{\mathbb C}^{v}}\ominus W$, so $\langle\psi({{x}_{1}},{{y}_{1}})|{{\psi}_{1}}({{x}_{2}},{{y}_{2}})\rangle=0$, thus   Eq. (\ref{2.13}) holds. Similarly, we can prove the case of  $\langle\psi({{x}_{1}},{{y}_{1}})|{{\psi}_{1}}({{x}_{2}},{{y}_{2}})\rangle\ne0$. From the above,   Eq. (\ref{2.13}) holds.
\qed

\section{Existence}

In this section, we mainly discuss the existence of non-classical idempotent QLS$(v)$s,
non-classical $t$-MOQLS$(v)$s for $t=2,3$, and non-classical SOQLS$(v)$s   by using the   constructions given in Section 2.

\subsection{The existence of non-classical idempotent QLSs}
\label{02}

 For an idempotent QLS$(v)$ $\Phi$, when the diagonal elements are non-computational basis of ${\mathbb{C}}^{v}$, we can use a unitary matrix to act on $\Phi$ so that its  diagonal elements are the computational basis of ${\mathbb{C}}^{v}$. Thus,  we only need to consider the idempotent QLS$(v)$ where the diagonal elements are the computational basis of ${\mathbb{C}}^{v}$.
In addition, for a unit vector  $|\psi\rangle\in {\mathbb{C}}^{v}$, we can always let $|\psi\rangle=a_0|0\rangle+a_1|1\rangle+ \cdots+a_{v-1}|v-1\rangle$, where the coefficients $a_i\in\mathbb{C},$ $i=0,1,\ldots,n-1$, are not all 0.

 Next, We first consider the existence of an idempotent QLS$(v)$. For $v=5$,  we first give the following lemma.

\begin{lemma}\label{IQLS5}
Let $\Phi$ be an idempotent QLS$(5)$ indexed by $[5]$,  where the diagonal elements of $\Phi$ form the computational basis of ${\mathbb{C}}^{5}$. If there exist three different elements $m, x, y \in [5]$ such that $|x\rangle$ and  $|y\rangle$ are  on the diagonal of $\Phi$, $|m\rangle$ is not on the diagonal of $\Phi$ but  $|m\rangle$ is in the same row as $|x\rangle$ and in the same column as $|y\rangle$,  then there exists an element  of $[5]\backslash\{x,y,m\}$ which  is in the same row as $|x\rangle$ and $|m\rangle$  but not in the $m$ column, and there exists an element  of $[5]\backslash\{x,y,m\}$ which  is in the same column as $|y\rangle$ and $|m\rangle$  but not in the $m$ row.
\end{lemma}

\Proof Without loss of generality,  let  the idempotent QLS$(5)$, $\Phi$ have the following form  and
$$\centering {\begin{array}{cc}
\setlength{\arraycolsep}{2.8 pt}
\begin{array}{lc}
\mbox{}&
\begin{array}{cc c c c}0& \hspace*{10pt}    1& \hspace*{10pt}   2&  \hspace*{10pt}      3 &  \hspace*{10pt}   4 \end{array} \\
\begin{array}{c}0 \\1\\ 2\\ 3\\ 4\\ \end{array}&
\begin{array}{|c|c|c|c|c|}
\hline
          |x\rangle  &     |m\rangle &           &           &   \\
\hline
                    &     |y\rangle &           &           &    \\
\hline
                     &               & |z\rangle &           &    \\
\hline
                     &               &           & |u\rangle &     \\
\hline
                     &               &           &           & |v\rangle\\
\hline
\end{array}
\end{array}
\end{array}}$$
\vspace{0.1cm}

\noindent    $|m \rangle=|z \rangle$. Then for the row corresponding to  $|x\rangle,|m\rangle$, the elements $|{{\Phi}_{03}}\rangle\in span\{|y\rangle, |v\rangle\}$, $|{{\Phi}_{04}}\rangle\in span\{|y\rangle, |u\rangle\}$ and $\langle{{\Phi}_{03}}|{{\Phi}_{04}}\rangle=0$. Since $span\{|y\rangle, |u\rangle\}\cap span\{|y\rangle, |v\rangle\}=span\{|y\rangle\}$.
Therefore, $|{{\Phi}_{03}}\rangle\in span\{|v\rangle\}$ or $|{{\Phi}_{04}}\rangle\in span\{|u\rangle\}$. That is,  $|{{\Phi}_{03}}\rangle$ or $|{{\Phi}_{04}}\rangle$ is from the computational basis of ${\mathbb{C}}^{5}$. Similarly, for the column corresponding to  $|m\rangle, |y\rangle$, the conclusion is valid. \qed

\begin{theorem}\label{v=3,4,5 IQLS}
For $v\in\{3,4,5\}$, there  is no non-classical idempotent QLS$($$v$$)$.
\end{theorem}

\Proof $(1)$ $v=3$, there is no non-classical QLS$(3)$ from \cite{Paczos}, so there is no non-classical idempotent QLS$(3)$.

$(2)$ $v=4$, if there exists a non-classical idempotent QLS(4), $\Phi=(|{{\Phi}_{ij}}\rangle)$  as shown in the following table, $\{|{{\Phi}_{ii}}\rangle: i\in [4]\}$ forms the computational basis of ${{\mathbb C}^{4}}$,  and there is at least one element $|{{\Phi}_{ij}}\rangle$ with $i\ne j$ which is not from the computational basis of ${{\mathbb C}^{4}}$.  Without loss of generality, let $|{{\Phi}_{ij}}\rangle=|{{\Phi}_{01}}\rangle$.

\begin{table}[!ht]
    \centering
    \begin{tabular}{|c|c|c|c|}
    \hline
        ${\color{red}|0\rangle}$ & $\color{blue}|{{\Phi}_{01}}\rangle$ & $|{{\Phi}_{02}}\rangle$ & $|{{\Phi}_{03}}\rangle$\\
        \hline
        $|{{\Phi}_{10}}\rangle$ & ${\color{red}|1\rangle}$ & $|{{\Phi}_{12}}\rangle$ & $|{{\Phi}_{13}}\rangle$ \\
        \hline
        $|{{\Phi}_{20}}\rangle$ & $|{{\Phi}_{21}}\rangle$ & ${\color{red}|2\rangle}$ & $|{{\Phi}_{23}}\rangle$\\
        \hline
        $|{{\Phi}_{30}}\rangle$ & $|{{\Phi}_{31}}\rangle$ & $|{{\Phi}_{32}}\rangle$ & ${\color{red}|3\rangle}$\\
        \hline
    \end{tabular}
\end{table}

Let~$|{{\Phi}_{01}}\rangle={{a}_{0}}|2\rangle+{{a}_{1}}|3\rangle$, ${a}_{0}\ne 0$ and ${a}_{1}\ne 0$. Let $|{{\Phi}_{02}}\rangle={{a}_{2}}|1\rangle+{{a}_{3}}|3\rangle$, $|{{\Phi}_{03}}\rangle={{a}_{4}}|1\rangle+{{a}_{5}}|2\rangle$. Since $\langle|{{\Phi}_{01}}|{{\Phi}_{02}}\rangle=0$, $\langle|{{\Phi}_{01}}|{{\Phi}_{03}}\rangle=0$, then ${{a}_{1}}{{a}_{3}}=0, {{a}_{0}}{{a}_{5}}=0$. So ${{a}_{3}}=0, {{a}_{5}}=0$, then $|{{\Phi}_{02}}\rangle=|1\rangle, |{{\Phi}_{03}}\rangle=|1\rangle$, which is a
contradiction. Therefore, there is no  non-classical idempotent QLS$(4)$.

$(3)$ See Appendix A for the proof of $v=5$. \qed

\begin{theorem}
For $v\geq 6$, there exists a non-classical idempotent $QLS(v)$.
\end{theorem}
\Proof $(1)$ $v=6$, there exists a non-classical idempotent QLS$(6)$  as follows:
\begin{equation*}
    \centering
    \begin{tabular}{|c|c|c|c|c|c|}
    \hline
        ${\color{red}|0\rangle}$ & $|4\rangle$ & $|3\rangle$ & $|5\rangle$ & $|2\rangle$ & $|1\rangle$\\
        \hline
        $|3\rangle$ & ${\color{red}|1\rangle}$ & $|0\rangle$ & $|4\rangle$ & $|5\rangle$ & $|2\rangle$\\
        \hline
        $|5\rangle$ & $|-\rangle$ & ${\color{red}|2\rangle}$ & $|1\rangle$ & $|+\rangle$ & $|4\rangle$\\
        \hline
        $|4\rangle$ & $|2\rangle$ & $|5\rangle$ & ${\color{red}|3\rangle}$ & $|1\rangle$ & $|0\rangle$\\
        \hline
        $|2\rangle$ & $|5\rangle$ & $|1\rangle$ & $|0\rangle$ & ${\color{red}|4\rangle}$ & $|3\rangle$\\
        \hline
        $|1\rangle$ & $|+\rangle$ & $|4\rangle$ & $|2\rangle$ & $|-\rangle$ & ${\color{red}|5\rangle}$\\
        \hline
    \end{tabular}
\end{equation*}

\noindent where
~$|+\rangle=\frac{1}{\sqrt{2}}(|0\rangle+|3\rangle)$, ~$|-\rangle=\frac{1}{\sqrt{2}}(|0\rangle-|3\rangle)$.

$(2)$  $v=8$, there exists a non-classical idempotent QLS$(8)$  as follows:

\begin{equation*}\hspace{0cm}
\begin{small}
\setlength{\arraycolsep}{3.8 pt}
\begin{array}{|c|c|c|c|c|c|c|c|}
\hline
{\color{red}|0\rangle} & |6\rangle & |7\rangle^{+}  & |1\rangle^{-}& |2\rangle & |3\rangle & |4\rangle  & |5\rangle\\
\hline
|5\rangle & {\color{red}|1\rangle} & |4\rangle  & |6\rangle& |3\rangle & |2\rangle & |7\rangle  & |0\rangle\\
\hline
|6\rangle & |0\rangle & {\color{red}|2\rangle}  & |5\rangle& |7\rangle & |4\rangle & \frac{1}{\sqrt{5}}(2|3\rangle+i|1\rangle)  & \frac{1}{\sqrt{5}}(i|3\rangle+2|1\rangle)\\
\hline
|7\rangle & |2\rangle & |0\rangle  & {\color{red}|3\rangle}& \frac{1}{\sqrt{5}}(2|6\rangle+i|1\rangle) & \frac{1}{\sqrt{5}}(i|6\rangle+2|1\rangle) & |5\rangle  & |4\rangle\\
\hline
|1\rangle & |5\rangle & |3\rangle  & |0\rangle& {\color{red}|4\rangle} & |7\rangle & |2\rangle  & |6\rangle\\
\hline
|2\rangle & |7\rangle & |6\rangle  & |4\rangle& |0\rangle & {\color{red}|5\rangle} & \frac{1}{\sqrt{5}}(i|3\rangle+2|1\rangle)  & \frac{1}{\sqrt{5}}(2|3\rangle+i|1\rangle)\\
\hline
|3\rangle^{+} & |4\rangle^{-} & |1\rangle^{-}  & |7\rangle^{+}& |5\rangle & |0\rangle & {\color{red}|6\rangle}  & |2\rangle\\
\hline
|4\rangle^{-} & |3\rangle^{+} & |5\rangle  & |2\rangle& \frac{1}{\sqrt{5}}(i|6\rangle+2|1\rangle) & \frac{1}{\sqrt{5}}(2|6\rangle+i|1\rangle) & |0\rangle  & {\color{red}|7\rangle}\\
\hline
\end{array}
\end{small}
\end{equation*}

\noindent where
$|3\rangle^{+}=\frac{1}{\sqrt{2}}(|3\rangle+|4\rangle)$, $|4\rangle^{-}=\frac{1}{\sqrt{2}}(|3\rangle-|4\rangle)$;
$|7\rangle^{+}=\frac{1}{\sqrt{2}}(|1\rangle+|7\rangle)$, $|1\rangle^{-}=\frac{1}{\sqrt{2}}(|1\rangle-|7\rangle)$.\vspace{0.2cm}

$(3)$ When $v\geq 3$ and $v\ne 6, 8$, there exists a PBD$(v, \{3, 4, 5\})$ by Lemma \ref{PBD1}. Since there exists an idempotent LS$(v)$ for $v=3,4,5$ by (3) and (8) of Lemma \ref{md}, then there exists an idempotent  QLS$(v)$ on the computational basis of ${{\mathbb C}^{v}}$ for $v=3,4,5$. Therefore, there exists a non-classical idempotent QLS$(v)$  for   $v\geq 7$ and $v\ne 8$  from Construction  \ref{PBD2}.

From the above, there exists a non-classical idempotent QLS$(v)$ for $v\geq 6$.
\qed

\begin{theorem}\label{th3.6}
When $v\geq 6$, there exists a non-classical $2$-idempotent MOQLS$(v)$, except possibly for $v\in \{6-12, 14, 15, 18, 19, 23\}$.
\end{theorem}

\Proof There is no non-classical $2$-idempotent MOQLS$(v)$ for $v=3,4,5$ by Theorem  \ref{v=3,4,5 IQLS}.
For $v\in \{4,5,7,9,10,11\}$, $N(1^v) \geq 2 $ by (7) of Lemma  \ref{lem1.1}, then there exists a $2$-idempotent MOQLS$(v)$ based on the computational basis of $\mathbb{C}^{v}$. According to Lemma \ref{PBD1} and Construction \ref{con2.4},  when $v\notin \{6-12, 14, 15, 18, 19, 23, 26, 27, 30\}$,   there exists a non-classical 2-idempotent MOQLS($v$). Further,
according to Lemma \ref{lem2.6} and Construction \ref{con2.7}, when  $v\in\{26, 27, 30\}$, there exists a non-classical 2-idempotent MOQLS($v$). So the conclusion holds.
\qed

\subsection{The existence of non-classical MOQLSs}
\label{04}

In this subsection, we will give the existence of non-classical QLS$(v)$s and non-classical $t$-MOQLS($v$)s for $t=2,3$ via the filling in holes constructions. 

\begin{lemma}{\rm(\!\!\cite{Paczos})}\label{QLS0}
There exists no non-classical QLS$(2)$ and QLS$(3)$.
\end{lemma}

\begin{theorem}
For $v\geq 4$, there exists a non-classical QLS$(v)$.
\end{theorem}
\Proof
Combining with (1) of Lemma \ref{IOLS} and Construction \ref{IQOLS}, a non-classical QLS can be given by filling a QLS of order 2 based on the basis of $|+\rangle=\frac{1}{\sqrt{2}}(|i\rangle+|j\rangle)$ and $|-\rangle=\frac{1}{\sqrt{2}}(|i\rangle-|j\rangle)$, where $i,j$ are the elements in the hole set.\qed

\begin{theorem}\label{jielun1}
For $v\geq 4$, there exists a non-classical $2$-MOQLS$($$v$$)$, except possibly for  $v\in \{ 4,5,6,7\}$.
\end{theorem}
\Proof
There exists a non-classical 2-MOQLS($9$) from (2) of Lemma \ref{Md1}. When $v=8,10,11$, there exist non-classical $2$-MOQLS$(v)$s listed below.  When $v\geq 12$, there exists a 2-IMOLS($v;4$) from (2) of Lemma \ref{IOLS}. Moreover,  a 2-MOLS(4) exists from (3)  of Lemma \ref{lem1.1}, so there exists a non-classical 2-MOQLS($v$) by Construction \ref{OLS1}. In conclusion, $v\geq 8$, there exists a non-classical 2-MOQLS($v$). \vspace{0.1cm}

 When $v=8$,   a non-classical 2-MOQLS(8) is listed below:  \vspace{0.2cm}
\begin{equation*}\hspace{0cm}
\begin{small}
\setlength{\arraycolsep}{3.8 pt}
\begin{array}{|c|c|c|c|c|c|c|c|}
\hline
|0\rangle^{+} & |1\rangle^{-} & |2\rangle  & |4\rangle& |3\rangle & |6\rangle & |7\rangle  & |5\rangle\\
\hline
|1\rangle^{-} & |0\rangle^{+} & |3\rangle  & |5\rangle& |2\rangle & |7\rangle & |6\rangle  & |4\rangle\\
\hline
|2\rangle^{+} & |3\rangle^{-} & |0\rangle  & |6\rangle& |1\rangle & |4\rangle & |5\rangle  & |7\rangle\\
\hline
|4\rangle & |5\rangle & |6\rangle  & |0\rangle& |7\rangle & |2\rangle^{+} & |3\rangle^{-}  & |1\rangle\\
\hline
|3\rangle^{-} & |2\rangle^{+} & |1\rangle  & |7\rangle& |0\rangle & |5\rangle & |4\rangle  & |6\rangle\\
\hline
|6\rangle & |7\rangle & |4\rangle  & |2\rangle& |5\rangle & |0\rangle^{+} & |1\rangle^{-}  & |3\rangle\\
\hline
|7\rangle & |6\rangle & |5\rangle  & |3\rangle& |4\rangle & |1\rangle^{-} & |0\rangle^{+}  & |2\rangle\\
\hline
|5\rangle & |4\rangle & |7\rangle  & |1\rangle& |6\rangle & |3\rangle^{-} & |2\rangle^{+}  & |0\rangle\\
\hline
\end{array}\end{small}~~~~
\begin{small}\begin{array}{|c|c|c|c|c|c|c|c|}
\hline
|0\rangle & |2\rangle & |4\rangle  & |3\rangle& |6\rangle & |7\rangle^{+} & |5\rangle^{-}  & |1\rangle\\
\hline
|1\rangle & |3\rangle & |5\rangle  & |2\rangle& |7\rangle & |6\rangle & |4\rangle  & |0\rangle\\
\hline
|2\rangle & |0\rangle & |6\rangle  & |1\rangle& |4\rangle & |5\rangle^{-} & |7\rangle^{+}  & |3\rangle\\
\hline
|4\rangle & |6\rangle & |0\rangle  & |7\rangle& |2\rangle & |3\rangle & |1\rangle  & |5\rangle\\
\hline
|3\rangle & |1\rangle & |7\rangle  & |0\rangle& |5\rangle & |4\rangle & |6\rangle  & |2\rangle\\
\hline
|6\rangle & |4\rangle & |2\rangle  & |5\rangle& |0\rangle & |1\rangle & |3\rangle  & |7\rangle\\
\hline
|7\rangle^{+} & |5\rangle^{-} & |3\rangle  & |4\rangle& |1\rangle & |0\rangle & |2\rangle  & |6\rangle\\
\hline
|5\rangle^{-} & |7\rangle^{+} & |1\rangle  & |6\rangle& |3\rangle & |2\rangle & |0\rangle  & |4\rangle\\
\hline
\end{array}
\end{small}
\end{equation*}

\noindent where $|0\rangle^{+}=\frac{1}{\sqrt{2}}(|0\rangle+|1\rangle)$,  $|1\rangle^{-}=\frac{1}{\sqrt{2}}(|0\rangle-|1\rangle)$;
$|2\rangle^{+}=\frac{1}{\sqrt{2}}(|2\rangle+|3\rangle)$, $|3\rangle^{-}=\frac{1}{\sqrt{2}}(|2\rangle-|3\rangle)$;
$|7\rangle^{+}=\frac{1}{\sqrt{2}}(|5\rangle+|7\rangle)$, $|5\rangle^{-}=\frac{1}{\sqrt{2}}(|5\rangle-|7\rangle)$.\vspace{0.2cm}

When $v=10,11$, there is an ISOLS(10; 3) and an ISOLS(11; 3) \cite{Denes,Zhu}.
\noindent\begin{equation*}\hspace{0cm}
\setlength{\arraycolsep}{6.3 pt}
\begin{small}\begin{array}{|c|c|c|c|c|c|c|c|c|c|}
\hline
0 & 7 & 8  & 9& 1 & 3 & 5  & 2& 4 &6 \\
\hline
6 & 1 & 7  & 8& 9 & 2 & 4  & 3& 5 &0 \\
\hline
5 & 0 & 2  & 7& 8 & 9 & 3  & 4& 6 &1 \\
\hline
4 & 6 & 1  & 3& 7 & 8 & 9  & 5& 0 &2 \\
\hline
9 & 5 & 0  & 2& 4 & 7 & 8  & 6& 1 &3 \\
\hline
8 & 9 & 6  & 1& 3 & 5 & 7  & 0& 2 &4 \\
\hline
7 & 8 & 9  & 0& 2 & 4 & 6  & 1& 3 &5 \\
\hline
1 & 2 & 3  & 4& 5 & 6 & 0  &  &   &  \\
\hline
2 & 3 & 4  & 5& 6 & 0 & 1  &  &  & \\
\hline
3 & 4 & 5  & 6& 0 & 1 & 2  &  &  & \\
\hline
\end{array}~~~~
\setlength{\arraycolsep}{3.4 pt}
\setlength{\extrarowheight}{1.2pt} 
{\small
\begin{array}{|c|c|c|c|c|c|c|c|c|c|c|}
\hline
0 & 8 & 1  & 9& 7 & 2 & 10  & 3& 5 &4 &6 \\
\hline
4 & 1 & 8  & 2& 9 & 0 & 3  & 10& 6 &5 &7 \\
\hline
10 & 5 & 2 & 8& 3 & 9 & 1  & 4& 7 &6  &0 \\
\hline
5 & 10 & 6  & 3& 8 & 4 & 9  &2& 0 &7  &1 \\
\hline
3 & 6 & 10  & 7& 4 & 8 & 5  & 9& 1 &0 &2 \\
\hline
9 & 4 & 7  & 10& 0 & 5 & 8  & 6& 2 &1 &3 \\
\hline
7 & 9 & 5  & 0& 10 & 1 & 6  & 8& 3 &2 &4 \\
\hline
8 & 0 & 9  & 6& 1 & 10 & 2  & 7& 4 &3 &5 \\
\hline
6 & 7 & 0  & 1& 2 & 3 & 4  &  5&  & & \\
\hline
2 & 3 & 4  & 5& 6 & 7 & 0  &  1&  & & \\
\hline
1 & 2 & 3  & 4& 5 & 6 & 7  &  0&  & & \\
\hline
\end{array}}\end{small}
\end{equation*}
\hspace{2.2cm}  An ISOLS(10; 3). \hspace{4.5cm}  An ISOLS(11; 3).\vspace{0.1cm}

So we have a non-classical 2-MOQLS(10) and a non-classical 2-MOQLS(11) by Construction \ref{OLS1}.
\begin{equation*}
\setlength{\arraycolsep}{1 pt}\hspace{0cm}
\setlength{\extrarowheight}{-1pt} 
\begin{array}{|c|c|c|c|c|c|c|c|c|c|}
\hline
|0\rangle & |7\rangle & |8\rangle  & |9\rangle& |1\rangle & |3\rangle & |5\rangle  & |2\rangle& |4\rangle &|6\rangle \\
\hline
|6\rangle & |1\rangle & |7\rangle & |8\rangle& |9\rangle & |2\rangle & |4\rangle  & |3\rangle& |5\rangle &|0\rangle \\
\hline
|5\rangle & |0\rangle & |2\rangle  & |7\rangle& |8\rangle & |9\rangle & |3\rangle  & |4\rangle& |6\rangle &|1\rangle \\
\hline
|4\rangle & |6\rangle & |1\rangle  & |3\rangle& |7\rangle & |8\rangle & |9\rangle  & |5\rangle& |0\rangle &|2\rangle \\
\hline
|9\rangle & |5\rangle & |0\rangle & |2\rangle& |4\rangle & |7\rangle & |8\rangle  & |6\rangle& |1\rangle &|3\rangle \\
\hline
|8\rangle & |9\rangle & |6\rangle  & |1\rangle& |3\rangle & |5\rangle & |7\rangle  & |0\rangle& |2\rangle &|4\rangle \\
\hline
|7\rangle & |8\rangle & |9\rangle  & |0\rangle& |2\rangle & |4\rangle & |6\rangle  & |1\rangle& |3\rangle &|5\rangle \\
\hline
|1\rangle & |2\rangle & |3\rangle  & |4\rangle& |5\rangle & |6\rangle & |0\rangle  & \textcolor[rgb]{1.00,0.00,0.00}{|7\rangle^{'}}& \textcolor[rgb]{1.00,0.00,0.00}{| 8\rangle^{'}} &\textcolor[rgb]{1.00,0.00,0.00}{|9\rangle^{'}} \\
\hline
|2\rangle & |3\rangle & |4\rangle  & |5\rangle& |6\rangle & |0\rangle & |1\rangle  &\textcolor[rgb]{1.00,0.00,0.00}{|8\rangle^{'}}& \textcolor[rgb]{1.00,0.00,0.00}{|9\rangle^{'}} &\textcolor[rgb]{1.00,0.00,0.00}{|7\rangle^{'}}\\
\hline
|3\rangle & |4\rangle & |5\rangle  & |6\rangle& |0\rangle & |1\rangle & |2\rangle  & \textcolor[rgb]{1.00,0.00,0.00}{|9\rangle^{'}}&\textcolor[rgb]{1.00,0.00,0.00}{ |7 \rangle^{'}} &\textcolor[rgb]{1.00,0.00,0.00}{|8 \rangle^{'}}\\
\hline
\end{array}~~
\begin{array}{|c|c|c|c|c|c|c|c|c|c|}
\hline
|0\rangle & |6\rangle & |5\rangle  & |4\rangle& |9\rangle & |8\rangle & |7\rangle  & |1\rangle& |2\rangle &|3\rangle \\
\hline
|7\rangle & |1\rangle & |0\rangle & |6\rangle& |5\rangle & |9\rangle & |8\rangle  & |2\rangle& |3\rangle &|4\rangle \\
\hline
|8\rangle & |7\rangle & |2\rangle  & |1\rangle& |0\rangle & |6\rangle & |9\rangle  & |3\rangle& |4\rangle &|5\rangle \\
\hline
|9\rangle & |8\rangle & |7\rangle  & |3\rangle& |2\rangle & |1\rangle & |0\rangle  & |4\rangle& |5\rangle &|6\rangle \\
\hline
|1\rangle & |9\rangle & |8\rangle & |7\rangle& |4\rangle & |3\rangle & |2\rangle  & |5\rangle& |6\rangle &|0\rangle \\
\hline
|3\rangle & |2\rangle & |9\rangle  & |8\rangle& |7\rangle & |5\rangle & |4\rangle  & |6\rangle& |0\rangle &|1\rangle \\
\hline
|5\rangle & |4\rangle & |3\rangle  & |9\rangle& |8\rangle & |7\rangle & |6\rangle  & |0\rangle& |1\rangle &|2\rangle \\
\hline
|2\rangle & |3\rangle & |4\rangle  & |5\rangle& |6\rangle & |0\rangle & |1\rangle  & \textcolor[rgb]{1.00,0.00,0.00}{|7\rangle^{'}}& \textcolor[rgb]{1.00,0.00,0.00}{| 8\rangle^{'}} &\textcolor[rgb]{1.00,0.00,0.00}{|9\rangle^{'}} \\
\hline
|4\rangle & |5\rangle & |6\rangle  & |0\rangle& |1\rangle & |2\rangle & |3\rangle  & \textcolor[rgb]{1.00,0.00,0.00}{|9\rangle^{'}}&\textcolor[rgb]{1.00,0.00,0.00}{ |7 \rangle^{'}} &\textcolor[rgb]{1.00,0.00,0.00}{|8 \rangle^{'}}\\
\hline
|6\rangle & |0\rangle & |1\rangle  & |2\rangle& |3\rangle & |4\rangle & |5\rangle  & \textcolor[rgb]{1.00,0.00,0.00}{|8\rangle^{'}}& \textcolor[rgb]{1.00,0.00,0.00}{|9\rangle^{'}} &\textcolor[rgb]{1.00,0.00,0.00}{|7\rangle^{'}}\\
\hline
\end{array}
\end{equation*}
\hspace{5cm}  A non-classical 2-MOQLS(10).
\begin{equation*}
\setlength{\arraycolsep}{0 pt}\hspace{0cm}
\begin{array}{|c|c|c|c|c|c|c|c|c|c|c|}
\hline
|0\rangle & |8\rangle & |1\rangle  & |9\rangle& |7\rangle & |2\rangle & |10\rangle  & |3\rangle& |5\rangle &|4\rangle &|6\rangle\\
\hline
|4\rangle & |1\rangle & |8\rangle  & |2\rangle& |9\rangle & |0\rangle & |3\rangle  & |10\rangle& |6\rangle &|5\rangle &|7\rangle\\
\hline
|10\rangle & |5\rangle & |2\rangle  & |8\rangle& |3\rangle & |9\rangle & |1\rangle  & |4\rangle& |7\rangle &|6\rangle &|0\rangle\\
\hline
|5\rangle & |10\rangle & |6\rangle  & |3\rangle& |8\rangle & |4\rangle & |9\rangle  & |2\rangle& |0\rangle &|7\rangle &|1\rangle\\
\hline
|3\rangle & |6\rangle & |10\rangle  & |7\rangle& |4\rangle & |8\rangle & |5\rangle  & |9\rangle& |1\rangle &|0\rangle &|2\rangle\\
\hline
|9\rangle & |4\rangle & |7\rangle  & |10\rangle& |0\rangle & |5\rangle & |8\rangle  & |6\rangle& |2\rangle &|1\rangle &|3\rangle\\
\hline
|7\rangle & |9\rangle & |5\rangle  & |0\rangle& |10\rangle & |1\rangle & |6\rangle  & |8\rangle& |3\rangle &|2\rangle &|4\rangle\\
\hline
|8\rangle & |0\rangle & |9\rangle  & |6\rangle& |1\rangle & |10\rangle & |2\rangle  &|7\rangle& |4\rangle& | 3\rangle &|5\rangle \\
\hline
|6\rangle & |7\rangle & |0\rangle  & |1\rangle& |2\rangle & |3\rangle & |4\rangle  &|5\rangle& \textcolor[rgb]{1.00,0.00,0.00}{|8\rangle^{''}}& \textcolor[rgb]{1.00,0.00,0.00}{| 9\rangle^{''}} &\textcolor[rgb]{1.00,0.00,0.00}{|10\rangle^{''}} \\
\hline
|2\rangle & |3\rangle & |4\rangle  & |5\rangle& |6\rangle & |7\rangle & |0\rangle  & |1\rangle& \textcolor[rgb]{1.00,0.00,0.00}{|9\rangle^{''}}& \textcolor[rgb]{1.00,0.00,0.00}{| 10\rangle^{''}} &\textcolor[rgb]{1.00,0.00,0.00}{|8\rangle^{''}} \\
\hline
|1\rangle & |2\rangle & |3\rangle  & |4\rangle& |5\rangle & |6\rangle & |7\rangle  &|0\rangle& \textcolor[rgb]{1.00,0.00,0.00}{|10\rangle^{''}}& \textcolor[rgb]{1.00,0.00,0.00}{| 8\rangle^{''}} &\textcolor[rgb]{1.00,0.00,0.00}{|9\rangle^{''}} \\
\hline
\end{array}~~
\begin{array}{|c|c|c|c|c|c|c|c|c|c|c|}
\hline
|0\rangle & |4\rangle & |10\rangle  & |5\rangle& |3\rangle & |9\rangle & |7\rangle  & |8\rangle& |6\rangle &|2\rangle &|1\rangle\\
\hline
|8\rangle & |1\rangle & |5\rangle  & |10\rangle& |6\rangle & |4\rangle & |9\rangle  & |0\rangle& |7\rangle &|3\rangle &|2\rangle\\
\hline
|1\rangle & |8\rangle & |2\rangle  & |6\rangle& |10\rangle & |7\rangle & |5\rangle  & |9\rangle& |0\rangle &|4\rangle &|3\rangle\\
\hline
|9\rangle & |2\rangle & |8\rangle  & |3\rangle& |7\rangle & |10\rangle & |0\rangle  & |6\rangle& |1\rangle &|5\rangle &|4\rangle\\
\hline
|7\rangle & |9\rangle & |3\rangle  & |8\rangle& |4\rangle & |0\rangle & |10\rangle  & |1\rangle& |2\rangle &|6\rangle &|5\rangle\\
\hline
|2\rangle & |0\rangle & |9\rangle  & |4\rangle& |8\rangle & |5\rangle & |1\rangle  & |10\rangle& |3\rangle &|7\rangle &|6\rangle\\
\hline
|10\rangle & |3\rangle & |1\rangle  & |9\rangle& |5\rangle & |8\rangle & |6\rangle  & |2\rangle& |4\rangle &|0\rangle &|7\rangle\\
\hline
|3\rangle & |10\rangle & |4\rangle  & |2\rangle& |9\rangle & |6\rangle & |8\rangle  &|7\rangle& |5\rangle& | 1\rangle &|0\rangle \\
\hline
|5\rangle & |6\rangle & |7\rangle  & |0\rangle& |1\rangle & |2\rangle & |3\rangle  &|4\rangle& \textcolor[rgb]{1.00,0.00,0.00}{|8\rangle^{''}}& \textcolor[rgb]{1.00,0.00,0.00}{| 9\rangle^{''}} &\textcolor[rgb]{1.00,0.00,0.00}{|10\rangle^{''}} \\
\hline
|4\rangle & |5\rangle & |6\rangle  & |7\rangle& |0\rangle & |1\rangle & |2\rangle  & |3\rangle& \textcolor[rgb]{1.00,0.00,0.00}{|10\rangle^{''}}& \textcolor[rgb]{1.00,0.00,0.00}{| 8\rangle^{''}} &\textcolor[rgb]{1.00,0.00,0.00}{|9\rangle^{''}} \\
\hline
|6\rangle & |7\rangle & |0\rangle  & |1\rangle& |2\rangle & |3\rangle & |4\rangle  &|5\rangle& \textcolor[rgb]{1.00,0.00,0.00}{|9\rangle^{''}}& \textcolor[rgb]{1.00,0.00,0.00}{| 10\rangle^{''}} &\textcolor[rgb]{1.00,0.00,0.00}{|8\rangle^{''}} \\
\hline
\end{array}
\end{equation*}
\hspace{5cm}  A non-classical 2-MOQLS(11).  \vspace{0.1cm}

\noindent where   $|i\rangle^{'}=U_0 |i\rangle$ for $i=7,8,9$,  $|j\rangle^{''}=U_1 |j\rangle$ for $j=8,9,10$, $U_0$ and $U_1$ are defined as following:
\hspace{5cm}\begin{equation}
\setlength{\arraycolsep}{0.7 pt}
U_0=\left(
    \begin{array}{cc}
    I_7&\\
    &V
 \end{array}
        \right),~~~~~~~U_1=\left(
    \begin{array}{cc}
    I_8&\\
    &V
 \end{array}
        \right),
\end{equation}

 \noindent $I_n$ is the identity matrix of order $n$, and

\begin{equation}
\setlength{\arraycolsep}{0.7 pt}
 V=\frac{1}{\sqrt{3}}\left(
       \begin{array}{ccc}
1 & 1& 1\\
1 & e^{\frac{2\pi \sqrt{-1}}{3}}& e^{\frac{-2\pi \sqrt{-1}}{3}}\\
1 & e^{\frac{-2\pi \sqrt{-1}}{3}}& e^{\frac{2\pi \sqrt{-1}}{3}}\\
\end{array}
        \right).
\end{equation}
\qed

\begin{theorem}\label{jielun1}
For $v\geq 4$, there exists a non-classical $3$-MOQLS$(v)$, except possibly for $v\in \{4,5,\dots,15\}$.
\end{theorem}
\Proof
When $v\geq 16$, there exists a 3-IMOLS($v;4$) by (3) of Lemma \ref{IOLS}. Combining with (4) of Lemma \ref{lem1.1} and Corollary \ref{IMOLS}, a non-classical 3-MOQLS($v$) can be obtained.
\qed

\subsection{The existence of non-classical SOQLSs}
\label{04}

\begin{theorem}\label{jielun1}
For $v\geq 13$, there exists a non-classical SOQLS$(v)$.
\end{theorem}
\Proof When $v\geq 13$, there exists an ISOLS($v;4$) by (4) of Lemma \ref{IOLS}.~According to Lemma \ref{lem1.1} and Construction \ref{ISOLS1}, a non-classical SOQLS($v$) can be obtained.
\qed

\section{Conclusions}
\label{conclusion}
$~\vspace{-0.5cm}$

 Latin squares play an important role in practical applications, especially some Latin squares with special properties, such as  SOLS, idempotent LS and so on. The QLS is a generalization of the LS. In analogy with the LS, we introduce the QLS, $t$-MOQLS, especially discuss the properties of idempotent QLS and SOQLS. By comparing some results obtained in this paper, we can see that the results of idempotent QLS and idempotent MOQLS are inferior to those of QLS and MOQLS, because idempotent QLS have stronger properties, which is consistent with our cognition. However, the cardinality of idempotent QLS constructed by PBD is larger than that of QLS. The cardinality of a Latin square is the number of its vectors different up to a phase, which is as the measurement of the difference between an LS and a QLS \cite{Paczos}. For QLSs with the same order, it is also meaningful to consider different cardinality, which is also an open problem raised in \cite{Paczos}.

\section*{Acknowledgements}
 
The authors would like to acknowledge Prof. Lie Zhu  from Suzhou University  for useful discussion on this paper.

\vspace{0.5cm}

{\centering\title{\bf\Large{ Appendix: The continuation of the proof in Theorem \ref{v=3,4,5 IQLS}.}}}\vspace{0.2cm}
\label{sec:AppA}

\Proof $(3)$ $v=5$, we prove it in terms of cardinality $c$ of a QLS, the   number of its vectors distinct up to a global phase.  Suppose  there exists a non-classical idempotent QLS$(5)$, $\Phi=(|{{\Phi}_{ij}}\rangle)$ as shown in the following:
\begin{equation*}
   \scalebox{1.1}{ \begin{tabular}{|c|c|c|c|c|}
    \hline
        ${\color{red}|0\rangle}$ & $|{{\Phi}_{01}}\rangle$ & $|{{\Phi}_{02}}\rangle$ & $|{{\Phi}_{03}}\rangle$ & $|{{\Phi}_{04}}\rangle$\\
        \hline
        $|{{\Phi}_{10}}\rangle$ & ${\color{red}|1\rangle}$ & $|{{\Phi}_{12}}\rangle$ & $|{{\Phi}_{13}}\rangle$ & $|{{\Phi}_{14}}\rangle$\\
        \hline
        $|{{\Phi}_{20}}\rangle$ & $|{{\Phi}_{21}}\rangle$ & ${\color{red}|2\rangle}$ & $|{{\Phi}_{23}}\rangle$ & $|{{\Phi}_{24}}\rangle$\\
        \hline
        $|{{\Phi}_{30}}\rangle$ & $|{{\Phi}_{31}}\rangle$ & $|{{\Phi}_{32}}\rangle$ & ${\color{red}|3\rangle}$ & $|{{\Phi}_{34}}\rangle$\\
        \hline
        $|{{\Phi}_{40}}\rangle$ & $|{{\Phi}_{41}}\rangle$ & $|{{\Phi}_{42}}\rangle$ & $|{{\Phi}_{43}}\rangle$  &${\color{red}|4\rangle}$ \\
        \hline
    \end{tabular}}
\end{equation*}
where ${|{{\Phi}_{ii}}\rangle}=|i\rangle, i=0,1,\ldots,4$~ form a computational basis of ${{\mathbb C}^{5}}$.\vspace{0.2cm}

\textbf{Case I.}  $5<c<25$~

In this case, according to the definition of cardinality, there must exist a non-diagonal element $|{{\Phi}_{ij}}\rangle \in \{|0\rangle,|1\rangle,\ldots,|4\rangle\}$. Without loss of generality,  let  $|{{\Phi}_{01}}\rangle=|2\rangle$.
 Next, we apply the conclusion of Lemma \ref{IQLS5} to find the contradictions.

First, we select three elements $|{{\Phi}_{00}}\rangle, |{{\Phi}_{01}}\rangle, |{{\Phi}_{11}}\rangle$. According to Lemma \ref{IQLS5}, with the row of $|{{\Phi}_{00}}\rangle=|0\rangle, |{{\Phi}_{01}}\rangle=|2\rangle$, we can confirm that $|{{\Phi}_{03}}\rangle=|4\rangle$ or $|{{\Phi}_{04}}\rangle=|3\rangle$. Similarly, with the column of $|{{\Phi}_{01}}\rangle=|2\rangle, |{{\Phi}_{11}}\rangle=|1\rangle$, we can confirm that  $|{{\Phi}_{31}}\rangle=|4\rangle$ or $|{{\Phi}_{41}}\rangle=|3\rangle$. In this way, we can get the following four cases:
\begin{equation*}
\centering 
\begin{minipage}{0.2\textwidth}
    \centering
    \scalebox{0.8}{\begin{tabular}{|p{0.4cm}|p{0.4cm}|p{0.4cm}|p{0.4cm}|p{0.4cm}|}
    \hline
        ${\color{red}|0\rangle}$ & $|2\rangle$ &  & $|4\rangle$ & \\
        \hline
          & ${\color{red}|1\rangle}$ &  &  & \\
        \hline
          &  & ${\color{red}|2\rangle}$ &  & \\
        \hline
          & $|4\rangle$ &  & ${\color{red}|3\rangle}$ & \\
        \hline
          &  &  &   &${\color{red}|4\rangle}$ \\
        \hline
 \end{tabular}}

\end{minipage}\hfill \hspace{0.6cm} 
\begin{minipage}{0.2\textwidth}
    \centering
    \scalebox{0.8}{\begin{tabular}{|p{0.4cm}|p{0.4cm}|p{0.4cm}|p{0.4cm}|p{0.4cm}|}
    \hline
        ${\color{red}|0\rangle}$ & $|2\rangle$ &  & $|4\rangle$ & \\
        \hline
          & ${\color{red}|1\rangle}$ &  &  & \\
        \hline
          &  & ${\color{red}|2\rangle}$ &  & \\
        \hline
          &  &  & ${\color{red}|3\rangle}$ & \\
        \hline
          & $|3\rangle$ &  &   &${\color{red}|4\rangle}$ \\
        \hline
    \end{tabular}}

\end{minipage}\hfill \hspace{0.6cm}
\begin{minipage}{0.2\textwidth}
    \centering
    \scalebox{0.8}{\begin{tabular}{|p{0.4cm}|p{0.4cm}|p{0.4cm}|p{0.4cm}|p{0.4cm}|}
    \hline
        ${\color{red}|0\rangle}$ & $|2\rangle$ &  &  & $|3\rangle$\\
        \hline
          & ${\color{red}|1\rangle}$ &  &  & \\
        \hline
          &  & ${\color{red}|2\rangle}$ &  & \\
        \hline
          & $|4\rangle$ &  & ${\color{red}|3\rangle}$ & \\
        \hline
          &  &  &   &${\color{red}|4\rangle}$ \\
        \hline
    \end{tabular}}

\end{minipage}\hfill \hspace{0.6cm}
\begin{minipage}{0.2\textwidth}
    \centering
    \scalebox{0.8}{\begin{tabular}{|p{0.4cm}|p{0.4cm}|p{0.4cm}|p{0.4cm}|p{0.4cm}|}
    \hline
        ${\color{red}|0\rangle}$ & $|2\rangle$ &  &  &$|3\rangle$ \\
        \hline
          & ${\color{red}|1\rangle}$ &  &  & \\
        \hline
          &  & ${\color{red}|2\rangle}$ &  & \\
        \hline
          &  &  & ${\color{red}|3\rangle}$ & \\
        \hline
          & $|3\rangle$ &  &   &${\color{red}|4\rangle}$ \\
        \hline
    \end{tabular}}

\end{minipage}
\end{equation*}

\hspace{1.1cm}  Case~1 \hspace{2.2cm}  Case~2 \hspace{2.4cm}  Case~3 \hspace{2.4cm}  Case~4

We use the method of proof by contradiction to determine the remaining elements in the first row and the second column.

For Case 1, we consider the elements $|{{\Phi}_{02}}\rangle$ and $|{{\Phi}_{04}}\rangle$, where  $|{{\Phi}_{02}}\rangle, |{{\Phi}_{04}}\rangle\in span\{|1\rangle, |3\rangle\}$. Suppose that
 $|{{\Phi}_{02}}\rangle,$ $ |{{\Phi}_{04}}\rangle$  are the entangled states of  $|1\rangle, |3\rangle$. According to the orthogonality of elements, we know that \vspace{0.1cm}

$|{{\Phi}_{32}}\rangle=a_0|0\rangle+a_1|1\rangle$$\rightarrow$ $|{{\Phi}_{32}}\rangle=|0\rangle$,

$|{{\Phi}_{12}}\rangle=a_0|0\rangle+a_1|3\rangle+a_2|4\rangle$$\rightarrow$  $|{{\Phi}_{12}}\rangle=|4\rangle$
$\rightarrow|{{\Phi}_{42}}\rangle=a_0|1\rangle+a_1|3\rangle$.

$|{{\Phi}_{34}}\rangle=a_0|1\rangle+a_1|2\rangle$$\rightarrow$ $|{{\Phi}_{34}}\rangle=|2\rangle$,

$|{{\Phi}_{14}}\rangle=a_0|0\rangle+a_1|2\rangle+a_2|3\rangle$$\rightarrow\langle{{\Phi}_{14}}|{{\Phi}_{34}}\rangle=0$
$\rightarrow|{{\Phi}_{24}}\rangle=a_0|1\rangle+a_1|3\rangle$.\vspace{0.1cm}

\noindent Further, there will be
 $|{{\Phi}_{30}}\rangle=|1\rangle, |{{\Phi}_{13}}\rangle=|2\rangle\rightarrow |{{\Phi}_{10}}\rangle=|3\rangle\rightarrow |{{\Phi}_{20}}\rangle=|4\rangle\rightarrow |{{\Phi}_{40}}\rangle=|2\rangle\rightarrow |{{\Phi}_{21}}\rangle=|0\rangle\rightarrow |{{\Phi}_{23}}\rangle=|1\rangle$.  Then $\langle{{\Phi}_{23}}|{{\Phi}_{24}}\rangle\ne0$ with contradiction. So we have~$|{{\Phi}_{02}}\rangle$~and~$|{{\Phi}_{04}}\rangle$ can only be $|1\rangle$ or $|3\rangle$.

Similarly, we consider the elements  $|{{\Phi}_{21}}\rangle$ and $|{{\Phi}_{41}}\rangle$, where $|{{\Phi}_{21}}\rangle, |{{\Phi}_{41}}\rangle\in span\{|0\rangle, |3\rangle\}$. If $|{{\Phi}_{21}}\rangle, |{{\Phi}_{41}}\rangle$ are the entangled states of $|0\rangle, |3\rangle$, then   \vspace{0.1cm}

 $|{{\Phi}_{23}}\rangle=a_0|0\rangle+a_1|1\rangle$$\rightarrow$$|{{\Phi}_{23}}\rangle=|1\rangle$,

$|{{\Phi}_{20}}\rangle=a_0|1\rangle+a_1|3\rangle+a_2|4\rangle$$\rightarrow $$|{{\Phi}_{20}}\rangle=|4\rangle$
$\rightarrow|{{\Phi}_{24}}\rangle=a_0|0\rangle+a_1|3\rangle$.

$|{{\Phi}_{43}}\rangle=a_0|0\rangle+a_1|2\rangle$$\rightarrow$ $|{{\Phi}_{43}}\rangle=|2\rangle$,

$|{{\Phi}_{40}}\rangle=a_0|1\rangle+a_1|2\rangle+a_2|3\rangle$$\rightarrow$ $|{{\Phi}_{40}}\rangle=|1\rangle$
$\rightarrow|{{\Phi}_{42}}\rangle=a_0|0\rangle+a_1|3\rangle$.\vspace{0.1cm}

\noindent Further, there will be
 $|{{\Phi}_{13}}\rangle=|0\rangle, |{{\Phi}_{30}}\rangle=|2\rangle\rightarrow |{{\Phi}_{10}}\rangle=|3\rangle\rightarrow |{{\Phi}_{12}}\rangle=|4\rangle\rightarrow |{{\Phi}_{14}}\rangle=|2\rangle\rightarrow |{{\Phi}_{32}}\rangle=|1\rangle\rightarrow |{{\Phi}_{02}}\rangle=|3\rangle$.  Then $\langle{{\Phi}_{02}}|{{\Phi}_{42}}\rangle\ne0$, which is a contradiction. Hence, $|{{\Phi}_{21}}\rangle$~and~$|{{\Phi}_{41}}\rangle$ can only be $|0\rangle$ or $|3\rangle$.

Based on the above analysis,   we can divide Case 1 into the following four sub-cases for further discussion.
\begin{equation*}
\centering 
\begin{minipage}{0.2\textwidth}
    \centering
    \scalebox{0.8}{\begin{tabular}{|p{0.4cm}|p{0.4cm}|p{0.4cm}|p{0.4cm}|p{0.4cm}|}
    \hline
        ${\color{red}|0\rangle}$ & $|2\rangle$ & $|1\rangle$ & $|4\rangle$ & $|3\rangle$\\
        \hline
          & ${\color{red}|1\rangle}$ &  &  & \\
        \hline
          & $|0\rangle$ & ${\color{red}|2\rangle}$ &  & \\
        \hline
          & $|4\rangle$ &  & ${\color{red}|3\rangle}$ & \\
        \hline
          & $|3\rangle$ &  &   &${\color{red}|4\rangle}$ \\
        \hline
 \end{tabular}}

\end{minipage}\hfill \hspace{0.6cm} 
\begin{minipage}{0.2\textwidth}
    \centering
    \scalebox{0.8}{\begin{tabular}{|p{0.4cm}|p{0.4cm}|p{0.4cm}|p{0.4cm}|p{0.4cm}|}
    \hline
        ${\color{red}|0\rangle}$ & $|2\rangle$ & $|1\rangle$ & $|4\rangle$ & $|3\rangle$\\
        \hline
          & ${\color{red}|1\rangle}$ &  &  & \\
        \hline
          & $|3\rangle$ & ${\color{red}|2\rangle}$ &  & \\
        \hline
          & $|4\rangle$ &  & ${\color{red}|3\rangle}$ & \\
        \hline
          & $|0\rangle$ &  &   &${\color{red}|4\rangle}$ \\
        \hline
    \end{tabular}}

\end{minipage}\hfill \hspace{0.6cm}
\begin{minipage}{0.2\textwidth}
    \centering
    \scalebox{0.8}{\begin{tabular}{|p{0.4cm}|p{0.4cm}|p{0.4cm}|p{0.4cm}|p{0.4cm}|}
    \hline
        ${\color{red}|0\rangle}$ & $|2\rangle$ & $|3\rangle$ & $|4\rangle$ & $|1\rangle$\\
        \hline
          & ${\color{red}|1\rangle}$ &  &  & \\
        \hline
          & $|0\rangle$ & ${\color{red}|2\rangle}$ &  & \\
        \hline
          & $|4\rangle$ &  & ${\color{red}|3\rangle}$ & \\
        \hline
          & $|3\rangle$ &  &   &${\color{red}|4\rangle}$ \\
        \hline
    \end{tabular}}

\end{minipage}\hfill \hspace{0.6cm}
\begin{minipage}{0.2\textwidth}
    \centering
    \scalebox{0.8}{\begin{tabular}{|p{0.4cm}|p{0.4cm}|p{0.4cm}|p{0.4cm}|p{0.4cm}|}
    \hline
        ${\color{red}|0\rangle}$ & $|2\rangle$ & $|3\rangle$ & $|4\rangle$ & $|1\rangle$\\
        \hline
          & ${\color{red}|1\rangle}$ &  &  & \\
        \hline
          & $|3\rangle$ & ${\color{red}|2\rangle}$ &  & \\
        \hline
          & $|4\rangle$ &  & ${\color{red}|3\rangle}$ & \\
        \hline
          & $|0\rangle$ &  &   &${\color{red}|4\rangle}$ \\
        \hline
    \end{tabular}}

\end{minipage}
\end{equation*}
\hspace{1.6cm}  Case~1-1 \hspace{2.1cm}  Case~1-2 \hspace{2cm}  Case~1-3 \hspace{2cm}  Case~1-4

According to Lemma \ref{IQLS5}, we use the method of proof by contradiction and the orthogonality of elements to determine the remaining elements.\vspace{0.1cm}

(1) Case 1-1

Select~$|{{\Phi}_{11}}\rangle, |{{\Phi}_{21}}\rangle, |{{\Phi}_{22}}\rangle$, then we can  determine  $|{{\Phi}_{23}}\rangle$ or $|{{\Phi}_{24}}\rangle$  according to Lemma \ref{IQLS5}. Thus, there are two  cases:
\begin{equation*}
\centering 
\begin{minipage}{0.5\textwidth}
    \centering
    \scalebox{0.8}{\begin{tabular}{|p{0.4cm}|p{0.4cm}|p{0.4cm}|p{0.4cm}|p{0.4cm}|}
    \hline
        ${\color{red}|0\rangle}$ & $|2\rangle$ & $|1\rangle$ & $|4\rangle$ & $|3\rangle$\\
        \hline
          & ${\color{red}|1\rangle}$ &  &  & \\
        \hline
          & $|0\rangle$ & ${\color{red}|2\rangle}$ & $|4\rangle$ & \\
        \hline
          & $|4\rangle$ &  & ${\color{red}|3\rangle}$ & \\
        \hline
          & $|3\rangle$ &  &   &${\color{red}|4\rangle}$ \\
        \hline
 \end{tabular}}

\end{minipage}\hfill
\begin{minipage}{0.5\textwidth}
    \centering
    \scalebox{0.8}{\begin{tabular}{|p{0.4cm}|p{0.4cm}|p{0.4cm}|p{0.4cm}|p{0.4cm}|}
    \hline
        ${\color{red}|0\rangle}$ & $|2\rangle$ & $|1\rangle$ & $|4\rangle$ & $|3\rangle$\\
        \hline
          & ${\color{red}|1\rangle}$ &  &  & \\
        \hline
          & $|0\rangle$ & ${\color{red}|2\rangle}$ &  & $|3\rangle$\\
        \hline
          & $|4\rangle$ &  & ${\color{red}|3\rangle}$ & \\
        \hline
          & $|3\rangle$ &  &   &${\color{red}|4\rangle}$ \\
        \hline
    \end{tabular}}

\end{minipage}
\end{equation*}
\hspace{2.6cm} Case~1-1-1 \hspace{5.8cm} Case~1-1-2
\vspace{0.2cm}\par

Obviously, these two cases are in contradiction with the definition of the QLS. So it can not constitute an idempotent QLS(5).\vspace{0.1cm}

(2)  Case 1-2

Select $|{{\Phi}_{11}}\rangle, |{{\Phi}_{21}}\rangle, |{{\Phi}_{22}}\rangle$, then we can determine $|{{\Phi}_{20}}\rangle$ or $|{{\Phi}_{24}}\rangle$ according to Lemma \ref{IQLS5}.  There are two  cases:
\begin{equation*}
\centering
\begin{minipage}{0.5\textwidth}
    \centering
    \scalebox{0.8}{\begin{tabular}{|c|c|c|c|c|}
    \hline
        ${\color{red}|0\rangle}$ & $|2\rangle$ & $|1\rangle$ & $|4\rangle$ & $|3\rangle$\\
        \hline
          & ${\color{red}|1\rangle}$ &  &  & \\
        \hline
         $|4\rangle$ & $|3\rangle$ & ${\color{red}|2\rangle}$ &  & \\
        \hline
          & $|4\rangle$ &  & ${\color{red}|3\rangle}$ & \\
        \hline
          & $|0\rangle$ &  &   &${\color{red}|4\rangle}$ \\
        \hline
 \end{tabular}}

\end{minipage}\hfill 
\begin{minipage}{0.5\textwidth}
    \centering
    \scalebox{0.8}{\begin{tabular}{|c|c|c|c|c|}
    \hline
        ${\color{red}|0\rangle}$ & $|2\rangle$ & $|1\rangle$ & $|4\rangle$ & $|3\rangle$\\
        \hline
          & ${\color{red}|1\rangle}$ &  &  & \\
        \hline
          & $|3\rangle$ & ${\color{red}|2\rangle}$ &  & $|0\rangle$\\
        \hline
          & $|4\rangle$ &  & ${\color{red}|3\rangle}$ & \\
        \hline
          & $|0\rangle$ &  &   &${\color{red}|4\rangle}$ \\
        \hline
    \end{tabular}}

\end{minipage}
\end{equation*}
\hspace{2.6cm} Case~1-2-1 \hspace{5.8cm}  Case~1-2-2

1) Case 1-2-1

 Suppose $|{{\Phi}_{23}}\rangle, |{{\Phi}_{24}}\rangle$ are the entangled states of $|0\rangle, |1\rangle$. Because $|{{\Phi}_{13}}\rangle=a_0|0\rangle+a_1 |2\rangle$$\rightarrow$$|{{\Phi}_{13}}\rangle=|2\rangle$,$|{{\Phi}_{43}}\rangle=a_0|1\rangle+ a_1|2\rangle$$\rightarrow$$|{{\Phi}_{43}}\rangle=|2\rangle$, which is a contradiction. Hence ~$|{{\Phi}_{23}}\rangle=|0\rangle$~or~$|1\rangle$. Furthermore, there are two cases:
\begin{equation*}
\centering
\begin{minipage}{0.5\textwidth}
    \centering
    \scalebox{0.8}{\begin{tabular}{|c|c|c|c|c|}
    \hline
        ${\color{red}|0\rangle}$ & $|2\rangle$ & $|1\rangle$ & $|4\rangle$ & $|3\rangle$\\
        \hline
          & ${\color{red}|1\rangle}$ &  &  & \\
        \hline
         $|4\rangle$ & $|3\rangle$ & ${\color{red}|2\rangle}$ & $|0\rangle$ & $|1\rangle$\\
        \hline
          & $|4\rangle$ &  & ${\color{red}|3\rangle}$ & \\
        \hline
          & $|0\rangle$ &  &   &${\color{red}|4\rangle}$ \\
        \hline
 \end{tabular}}

\end{minipage}\hfill 
\begin{minipage}{0.5\textwidth}
    \centering
    \scalebox{0.8}{\begin{tabular}{|c|c|c|c|c|}
    \hline
        ${\color{red}|0\rangle}$ & $|2\rangle$ & $|1\rangle$ & $|4\rangle$ & $|3\rangle$\\
        \hline
          & ${\color{red}|1\rangle}$ &  &  & \\
        \hline
         $|4\rangle$ & $|3\rangle$ & ${\color{red}|2\rangle}$ & $|1\rangle$ & $|0\rangle$\\
        \hline
          & $|4\rangle$ &  & ${\color{red}|3\rangle}$ & \\
        \hline
          & $|0\rangle$ &  &   &${\color{red}|4\rangle}$ \\
        \hline
    \end{tabular}}

\end{minipage}
\end{equation*}
\hspace{2.6cm}  Case~1-2-1-1 \hspace{5.4cm}  Case~1-2-1-2

$\textcircled{1}$ Case 1-2-1-1

Since the elements in each row and column of $\Phi$ form a standard orthogonal basis of ${{\mathbb C}^{5}}$, then
$|{{\Phi}_{32}}\rangle=|0\rangle\rightarrow |{{\Phi}_{42}}\rangle=|3\rangle\rightarrow |{{\Phi}_{12}}\rangle=|4\rangle\rightarrow |{{\Phi}_{13}}\rangle=|2\rangle\rightarrow |{{\Phi}_{14}}\rangle=|0\rangle\rightarrow |{{\Phi}_{10}}\rangle=|3\rangle\rightarrow |{{\Phi}_{34}}\rangle=|2\rangle\rightarrow |{{\Phi}_{30}}\rangle=|1\rangle\rightarrow |{{\Phi}_{40}}\rangle=|2\rangle\rightarrow |{{\Phi}_{43}}\rangle=|1\rangle$. This precisely forms a classical idempotent QLS(5).\vspace{0.1cm}

$\textcircled{2}$ Case~1-2-1-2

Similar to $\textcircled{1}$, we have
$|{{\Phi}_{32}}\rangle=|0\rangle\rightarrow |{{\Phi}_{42}}\rangle=|3\rangle\rightarrow |{{\Phi}_{12}}\rangle=|4\rangle\rightarrow |{{\Phi}_{43}}\rangle=|2\rangle\rightarrow |{{\Phi}_{40}}\rangle=|1\rangle\rightarrow |{{\Phi}_{13}}\rangle=|0\rangle\rightarrow |{{\Phi}_{14}}\rangle=|2\rangle\rightarrow |{{\Phi}_{10}}\rangle=|3\rangle\rightarrow |{{\Phi}_{30}}\rangle=|2\rangle\rightarrow |{{\Phi}_{34}}\rangle=|1\rangle$. This forms a classical idempotent QLS(5).\vspace{0.1cm}

2) Case 1-2-2

Since the elements in each row and column of $\Phi$ form a standard orthogonal basis of ${{\mathbb C}^{5}}$, then
$|{{\Phi}_{32}}\rangle=|0\rangle\rightarrow |{{\Phi}_{42}}\rangle=|3\rangle\rightarrow |{{\Phi}_{12}}\rangle=|4\rangle\rightarrow |{{\Phi}_{23}}\rangle=|1\rangle\rightarrow |{{\Phi}_{20}}\rangle=|4\rangle\rightarrow |{{\Phi}_{43}}\rangle=|2\rangle\rightarrow |{{\Phi}_{40}}\rangle=|1\rangle\rightarrow |{{\Phi}_{30}}\rangle=|2\rangle\rightarrow |{{\Phi}_{10}}\rangle=|3\rangle\rightarrow |{{\Phi}_{13}}\rangle=|0\rangle\rightarrow |{{\Phi}_{14}}\rangle=|2\rangle\rightarrow |{{\Phi}_{34}}\rangle=|1\rangle$. This forms a classical idempotent QLS(5).

From the above discussion, there only exists a classical idempotent QLS(5) from Case 1-2.\vspace{0.1cm}

(3) Case 1-3

Select $|{{\Phi}_{11}}\rangle, |{{\Phi}_{21}}\rangle, |{{\Phi}_{22}}\rangle$, then we can determine  $|{{\Phi}_{23}}\rangle$ or $|{{\Phi}_{24}}\rangle$  according to Lemma \ref{IQLS5}. Further, there are two  cases:
\begin{equation*}
\centering 
\begin{minipage}{0.5\textwidth}
    \centering
    \scalebox{0.8}{\begin{tabular}{|c|c|c|c|c|}
    \hline
        ${\color{red}|0\rangle}$ & $|2\rangle$ & $|3\rangle$ & $|4\rangle$ & $|1\rangle$\\
        \hline
          & ${\color{red}|1\rangle}$ &  &  & \\
        \hline
          & $|0\rangle$ & ${\color{red}|2\rangle}$ & $|4\rangle$ & \\
        \hline
          & $|4\rangle$ &  & ${\color{red}|3\rangle}$ & \\
        \hline
          & $|3\rangle$ &  &   &${\color{red}|4\rangle}$ \\
        \hline
 \end{tabular}}

\end{minipage}\hfill 
\begin{minipage}{0.5\textwidth}
    \centering
    \scalebox{0.8}{\begin{tabular}{|c|c|c|c|c|}
    \hline
        ${\color{red}|0\rangle}$ & $|2\rangle$ & $|3\rangle$ & $|4\rangle$ & $|1\rangle$\\
        \hline
          & ${\color{red}|1\rangle}$ &  &  & \\
        \hline
          & $|0\rangle$ & ${\color{red}|2\rangle}$ &  & $|3\rangle$\\
        \hline
          & $|4\rangle$ &  & ${\color{red}|3\rangle}$ & \\
        \hline
          & $|3\rangle$ &  &   &${\color{red}|4\rangle}$ \\
        \hline
    \end{tabular}}

\end{minipage}
\end{equation*}
\hspace{2.6cm}  Case~1-3-1 \hspace{5.9cm}  Case~1-3-2

1) Case 1-3-1

Obviously, this case is contradictory to the definition of QLS, so it does not constitute an idempotent QLS(5).

2) Case 1-3-2

If $|{{\Phi}_{20}}\rangle, |{{\Phi}_{23}}\rangle$ are the entangled states of $|1\rangle, |4\rangle$, then $|{{\Phi}_{23}}\rangle=|1\rangle$, $|{{\Phi}_{20}}\rangle=|4\rangle$ because $|{{\Phi}_{03}}\rangle=|4\rangle, \langle{{\Phi}_{03}}|{{\Phi}_{23}}\rangle=0$. Select $|{{\Phi}_{11}}\rangle, |{{\Phi}_{31}}\rangle, |{{\Phi}_{33}}\rangle$, then we can determine $|{{\Phi}_{30}}\rangle$ or $|{{\Phi}_{32}}\rangle$ according to Lemma \ref{IQLS5}. There will exist two  cases. Similar to Case 1-2, there only exists an classical idempotent QLS(5).\vspace{0.1cm}

(4) Case 1-4

For a discussion similar to Case 1-2, there exist two classical idempotent QLS(5)s at this case.

From the above, we  know that Case 1 can only provide an classical idempotent QLS(5). \vspace{0.1cm}

For the  Cases 2-4, when we discuss them in the way of Case 1, it can be known that only  an classical idempotent QLS(5) exists. Hence, when~$5<c<25 $~, there only exists an classical idempotent QLS(5). \vspace{0.2cm}

\textbf{Case II.}  $c=25$~

According to Table 1, let

$|{{\Phi}_{01}}\rangle={{a}_{1}}|2\rangle+{{a}_{2}}|3\rangle+{{a}_{3}}|4\rangle$,~~~~~~~~~~~$|{a_1}|^{2}+|{a_2}|^{2}+|{a_3}|^{2}=1$;

$|{{\Phi}_{02}}\rangle={{a}_{4}}|1\rangle+{{a}_{5}}|3\rangle+{{a}_{6}}|4\rangle$,~~~~~~~~~~~$|{a_4}|^{2}+|{a_5}|^{2}+|{a_6}|^{2}=1$;

$|{{\Phi}_{03}}\rangle={{a}_{7}}|1\rangle+{{a}_{8}}|2\rangle+{{a}_{9}}|4\rangle$,~~~~~~~~~~~$|{a_7}|^{2}+|{a_8}|^{2}+|{a_9}|^{2}=1$;

$|{{\Phi}_{04}}\rangle={{a}_{10}}|1\rangle+{{a}_{11}}|2\rangle+{{a}_{12}}|3\rangle$,~~~~~~~~$|{a}_{10}|^{2}+|{a}_{11}|^{2}+|{a}_{12}|^{2}=1$.

Since $\Phi$ is a quantum Latin square,  then\vspace{0.1cm}

~$\langle|{{\Phi}_{01}}|{{\Phi}_{02}}\rangle=\overline{{{a }_{2}}}{{a }_{5}}+\overline{{{a }_{3}}}{{a }_{6}}=0$,~~~~~~~~~~${{a }_{6}}=-\frac{\overline{{{a }_{2}}}}{\overline{{{a }_{3}}}}{{a }_{5}}$;

~$\langle|{{\Phi}_{01}}|{{\Phi}_{03}}\rangle=\overline{{{a }_{1}}}{{a }_{8}}+\overline{{{a }_{3}}}{{a }_{9}}=0$,~~~~~~~~~~${{a }_{9}}=-\frac{\overline{{{a}_{1}}}}{\overline{{{a}_{3}}}}{{a}_{8}}=\frac{{{a}_{3}}\overline{{{a}_{4}}}}{{{a}_{2}}\overline{{{a }_{5}}}}{{a }_{7}}$;

~$\langle|{{\Phi}_{01}}|{{\Phi}_{04}}\rangle=\overline{{{a}_{1}}}{{a}_{11}}+\overline{{{a}_{2}}}{{a}_{12}}=0$,~~~~~~~~~~${{a }_{12}}=-\frac{\overline{{{a}_{1}}}}{\overline{{{a }_{2}}}}{{a}_{11}}=-\frac{\overline{{{a}_{4}}}}{\overline{{{a}_{5}}}}{{a}_{10}}$;

~$\langle|{{\Phi}_{02}}|{{\Phi}_{03}}\rangle=\overline{{{a}_{4}}}{{a}_{7}}+\overline{{{a}_{6}}}{{a}_{9}}=0$,~~~~~~~~~~~~${{a }_{8}}=-\frac{{{a}_{3}}\overline{{{a}_{3}}}\overline{{{a}_{4}}}}{\overline{{{a}_{1}}}{{a}_{2}}\overline{{{a}_{5}}}}{{a }_{7}}$;

~$\langle|{{\Phi}_{02}}|{{\Phi}_{04}}\rangle=\overline{{{a}_{4}}}{{a}_{10}}+\overline{{{a}_{5}}}{{a}_{12}}=0$,~~~~~~~~~~${{a }_{11}}=\frac{\overline{{{a}_{2}}}\overline{{{a}_{4}}}}{\overline{{{a }_{1}}}\overline{{{a}_{5}}}}{{a}_{10}}$;

~$\langle|{{\Phi}_{03}}|{{\Phi}_{04}}\rangle=\overline{{{a}_{7}}}{{a}_{10}}+\overline{{{a}_{8}}}{{a }_{11}}=0$;

\noindent  and

~$|{{\Phi}_{01}}\rangle={{a}_{1}}|2\rangle+{{a}_{2}}|3\rangle+{{a}_{3}}|4\rangle$,

~$|{{\Phi}_{02}}\rangle={{a}_{4}}|1\rangle+{{a}_{5}}|3\rangle-\frac{\overline{{{a}_{2}}}}{\overline{{{a}_{3}}}}{{a}_{5}}|4\rangle$,

~$|{{\Phi}_{03}}\rangle={{a}_{7}}|1\rangle-\frac{{{a}_{3}}\overline{{{a}_{3}}}\overline{{{a}_{4}}}}{\overline{{{a}_{1}}}{{a }_{2}}\overline{{{a}_{5}}}}{{a}_{7}}|2\rangle+\frac{{{a}_{3}}\overline{{{a}_{4}}}}{{{a}_{2}}\overline{{{a}_{5}}}}{{a}_{7}}|4\rangle$,

~$|{{\Phi}_{04}}\rangle={{a}_{10}}|1\rangle+\frac{\overline{{{a}_{2}}}\overline{{{a}_{4}}}}{\overline{{{a}_{1}}}\overline{{{a}_{5}}}}{{a }_{10}}|2\rangle-\frac{\overline{{{a}_{4}}}}{\overline{{{a}_{5}}}}{{a}_{10}}|3\rangle$.\vspace{0.1cm}

Let the elements in the first to fifth columns in $\Phi$ be as follows:\vspace{0.1cm}

$|{{\Phi}_{10}}\rangle={{f}_{1}}|2\rangle+{{f}_{2}}|3\rangle+{{f}_{3}}|4\rangle$,~~~~~~~~~~~$|{{\Phi}_{20}}\rangle={{f}_{4}}|1\rangle+{{f}_{5}}|3\rangle+{{f}_{6}}|4\rangle$,

$|{{\Phi}_{30}}\rangle={{f}_{7}}|1\rangle+{{f}_{8}}|2\rangle+{{f}_{9}}|4\rangle$,~~~~~~~~~~~$|{{\Phi}_{40}}\rangle={{f}_{10}}|1\rangle+{{f}_{11}}|2\rangle+{{f}_{12}}|3\rangle$;\par
\vspace{0.2cm}

$|{{\Phi}_{21}}\rangle={{b}_{1}}|0\rangle+{{b}_{2}}|3\rangle+{{b}_{3}}|4\rangle$,~~~~~~~~~~~~$|{{\Phi}_{31}}\rangle={{b}_{4}}|0\rangle+{{b}_{5}}|2\rangle+{{b}_{6}}|4\rangle$,

$|{{\Phi}_{41}}\rangle={{b}_{7}}|0\rangle+{{b}_{8}}|2\rangle+{{b}_{9}}|3\rangle$;\par
\vspace{0.2cm}

$|{{\Phi}_{12}}\rangle={{c}_{1}}|0\rangle+{{c}_{2}}|3\rangle+{{c}_{3}}|4\rangle$,~~~~~~~~~~~~$|{{\Phi}_{32}}\rangle={{c}_{4}}|0\rangle+{{c}_{5}}|1\rangle+{{c}_{6}}|4\rangle$,

$|{{\Phi}_{42}}\rangle={{c}_{7}}|0\rangle+{{c}_{8}}|1\rangle+{{c}_{9}}|3\rangle$;\par
\vspace{0.2cm}

$|{{\Phi}_{13}}\rangle={{d}_{1}}|0\rangle+{{d}_{2}}|2\rangle+{{d}_{3}}|4\rangle$,~~~~~~~~~~~$|{{\Phi}_{23}}\rangle={{d}_{4}}|0\rangle+{{d}_{5}}|1\rangle+{{d}_{6}}|4\rangle$,

$|{{\Phi}_{43}}\rangle={{d}_{7}}|0\rangle+{{d}_{8}}|1\rangle+{{d}_{9}}|2\rangle$;\par
\vspace{0.15cm}

$|{{\Phi}_{14}}\rangle={{e}_{1}}|0\rangle+{{e}_{2}}|2\rangle+{{e}_{3}}|3\rangle$,~~~~~~~~~~~~$|{{\Phi}_{24}}\rangle={{e}_{4}}|0\rangle+{{e}_{5}}|1\rangle+{{e}_{6}}|3\rangle$,

$|{{\Phi}_{34}}\rangle={{e}_{7}}|0\rangle+{{e}_{8}}|1\rangle+{{e}_{9}}|2\rangle$.\vspace{0.2cm}
 
\noindent According to the orthogonality of the elements in each column, we have

$|{{\Phi}_{10}}\rangle={{f}_{1}}|2\rangle+{{f}_{2}}|3\rangle+{{f}_{3}}|4\rangle$,
~~~~~~~~~~~~~~~~~~~$|{{\Phi}_{20}}\rangle={{f}_{4}}|1\rangle+{{f}_{5}}|3\rangle-\frac{\overline{{{f}_{2}}}}
{\overline{{{f}_{3}}}}{{f}_{5}}|4\rangle$,

$|{{\Phi}_{30}}\rangle={{f}_{7}}|1\rangle-\frac{{{f}_{3}}\overline{{{f}_{3}}{{f}_{4}}}}
{\overline{{{f}_{1}}}{{f}_{2}}\overline{{{f}_{5}}}}{{f}_{7}}|2\rangle+\frac{{{f}_{3}}\overline{{{f}_{4}}}}{{{f}_{2}}
\overline{{{f}_{5}}}}{{f}_{7}}|4\rangle$,
~~~~~~~$|{{\Phi}_{40}}\rangle={{f}_{10}}|1\rangle+\frac{\overline{{{f}_{2}}}\overline{{{f}_{4}}}}
{\overline{{{f}_{1}}}\overline{{{f}_{5}}}}{{f}_{10}}|2\rangle-\frac{\overline{{{f}_{4}}}}
{\overline{{{f}_{5}}}}{{f}_{10}}|3\rangle$;\par
\vspace{0.15cm}
$|{{\Phi}_{21}}\rangle={{b}_{1}}|0\rangle+{{b}_{2}}|3\rangle-\frac{\overline{{{a}_{2}}}}{\overline{{{a}_{3}}}}{{b}_{2}}|4\rangle$,
~~~~~~~~~~~~~~~~~$|{{\Phi}_{31}}\rangle={{b}_{4}}|0\rangle-\frac{{{a}_{3}}
\overline{{{a}_{3}}}\overline{{{b}_{1}}}}{\overline{{{a}_{1}}}{{a}_{2}}\overline{{{b}_{2}}}}{{b}_{4}}|2\rangle+\frac{{{a}_{3}}\overline{{{b}_{1}}}}{{{a}_{2}}\overline{{{b}_{2}}}}{{b}_{4}}|4\rangle$,

$|{{\Phi}_{41}}\rangle={{b}_{7}}|0\rangle+\frac{\overline{{{a}_{2}}}\overline{{{b}_{1}}}}{\overline{{{a}_{1}}}\overline{{{b}_{2}}}}{{b}_{7}}
|2\rangle-\frac{\overline{{{b}_{1}}}}{\overline{{{b}_{2}}}}{{b}_{7}}|3\rangle$;\par
\vspace{0.15cm}

$|{{\Phi}_{12}}\rangle={{c}_{1}}|0\rangle+{{c}_{2}}|3\rangle+\frac{{{a}_{3}}}{{{a}_{2}}}{{c}_{2}}|4\rangle$,
~~~~~~~~~~~~~~~~~$|{{\Phi}_{32}}\rangle={{c}_{4}}|0\rangle-\frac{{{a}_{2}}\overline{{{a}_{2}}}\overline{{{a}_{5}}}
\overline{{{c}_{1}}}}{{{a}_{3}}\overline{{{a}_{3}}}\overline{{{a}_{4}}}\overline{{{c}_{2}}}}{{c}_{4}}|1\rangle-
\frac{\overline{{{a}_{2}}}\overline{{{c}_{1}}}}{\overline{{{a}_{3}}}\overline{{{c}_{2}}}}{{c}_{4}}|4\rangle$,

$|{{\Phi}_{42}}\rangle={{c}_{7}}|0\rangle+\frac{\overline{{{a}_{5}}}\overline{{{c}_{1}}}}{\overline{{{a}_{4}}}
\overline{{{c}_{2}}}}{{c}_{7}}|1\rangle-\frac{\overline{{{c}_{1}}}}{\overline{{{c}_{2}}}}{{c}_{7}}|3\rangle$;\par
\vspace{0.15cm}

$|{{\Phi}_{13}}\rangle={{d}_{1}}|0\rangle+{{d}_{2}}|2\rangle+\frac{{{a}_{3}}}{{{a}_{1}}}{{d}_{2}}|4\rangle$,
~~~~~~~~~~~~~~~~$|{{\Phi}_{23}}\rangle={{d}_{4}}|0\rangle+\frac{\overline{{{a}_{1}}}{{a}_{4}}\overline{{{d}_{1}}}}{\overline{{{a}_{2}}}{{a}_{5}}
\overline{{{d}_{2}}}}{{d}_{4}}|1\rangle-\frac{\overline{{{a}_{1}}}\overline{{{d}_{1}}}}{\overline{{{a}_{3}}}
\overline{{{d}_{2}}}}{{d}_{4}}|4\rangle$,

$|{{\Phi}_{43}}\rangle={{d}_{7}}|0\rangle-\frac{{{a}_{3}}\overline{{{a}_{3}}}{{a}_{4}}\overline{{{d}_{1}}}}{{{a}_{1}}
\overline{{{a}_{2}}}{{a}_{5}}\overline{{{d}_{2}}}}{{d}_{7}}|1\rangle-\frac{\overline{{{d}_{1}}}}{\overline{{{d}_{2}}}}{{d}_{7}}
|2\rangle$;\par
\vspace{0.15cm}

$|{{\Phi}_{14}}\rangle={{e}_{1}}|0\rangle+{{e}_{2}}|2\rangle+\frac{{{a}_{2}}}{{{a}_{1}}}{{e}_{2}}|3\rangle$,
~~~~~~~~~~~~~~~~~$|{{\Phi}_{24}}\rangle={{e}_{4}}|0\rangle-\frac{\overline{{{a}_{1}}}{{a}_{4}}\overline{{{e}_{1}}}}{\overline{{{a}_{2}}}{{a}_{5}}
\overline{{{e}_{2}}}}{{e}_{4}}|1\rangle-\frac{\overline{{{a}_{1}}}\overline{{{e}_{1}}}}{\overline{{{a}_{2}}}
\overline{{{e}_{2}}}}{{e}_{4}}|3\rangle$,

$|{{\Phi}_{34}}\rangle={{e}_{7}}|0\rangle+\frac{{{a}_{2}}{{a}_{4}}\overline{{{e}_{1}}}}{{{a}_{1}}{{a}_{5}}\overline{{{e}_{2}}}}
{{e}_{7}}|1\rangle-\frac{\overline{{{e}_{1}}}}{\overline{{{e}_{2}}}}{{e}_{7}}|2\rangle$.\vspace{0.2cm}

Thus, we can obtain a quantum Latin square as shown in Table \ref{000}.
\begin{sidewaystable}[thp]
\renewcommand\arraystretch{1.5}
\caption{When $c=25$} \label{000}\vspace{0.4cm}
{\footnotesize{
\begin{tabular}{|c|c|c|c|c|}
\hline
  ${\color{red}|0\rangle}$ & ${{a}_{1}}|2\rangle+{{a}_{2}}|3\rangle+{{a}_{3}}|4\rangle$ & ${{a}_{4}}|1\rangle+{{a}_{5}}|3\rangle-\frac{\overline{{{a}_{2}}}}{\overline{{{a}_{3}}}}{{a}_{5}}|4\rangle$ & ${{a}_{7}}|1\rangle-\frac{{{a}_{3}}\overline{{{a}_{3}}}\overline{{{a}_{4}}}}{\overline{{{a}_{1}}}{{a}_{2}}\overline{{{a}_{5}}}}{{a }_{7}}|2\rangle+\frac{{{a}_{3}}\overline{{{a}_{4}}}}{{{a}_{2}}\overline{{{a}_{5}}}}{{a}_{7}}|4\rangle$ & ${{a}_{10}}|1\rangle+\frac{\overline{{{a}_{2}}}\overline{{{a}_{4}}}}{\overline{{{a}_{1}}}\overline{{{a}_{5}}}}{{a }_{10}}|2\rangle-\frac{\overline{{{a}_{4}}}}{\overline{{{a}_{5}}}}{{a}_{10}}|3\rangle$\\
  \hline
  ${{f}_{1}}|2\rangle+{{f}_{2}}|3\rangle+{{f}_{3}}|4\rangle$ & ${\color{red}|1\rangle}$ & ${{c}_{1}}|0\rangle+{{c}_{2}}|3\rangle+\frac{{{a}_{3}}}{{{a}_{2}}}{{c}_{2}}|4\rangle$ & ${{d}_{1}}|0\rangle+{{d}_{2}}|2\rangle+\frac{{{a}_{3}}}{{{a}_{1}}}{{d}_{2}}|4\rangle$ & ${{e}_{1}}|0\rangle+{{e}_{2}}|2\rangle+\frac{{{a}_{2}}}{{{a}_{1}}}{{e}_{2}}|3\rangle$\\
  \hline
  ${{f}_{4}}|1\rangle+{{f}_{5}}|3\rangle-\frac{\overline{{{f}_{2}}}}{\overline{{{f}_{3}}}}{{f}_{5}}|4\rangle$ & ${{b}_{1}}|0\rangle+{{b}_{2}}|3\rangle-\frac{\overline{{{a}_{2}}}}{\overline{{{a}_{3}}}}{{b}_{2}}|4\rangle$ & ${\color{red}|2\rangle}$ & ${{d}_{4}}|0\rangle+\frac{\overline{{{a}_{1}}}{{a}_{4}}\overline{{{d}_{1}}}}
  {\overline{{{a}_{2}}}{{a}_{5}}\overline{{{d}_{2}}}}{{d}_{4}}|1\rangle-\frac{\overline{{{a}_{1}}}
  \overline{{{d}_{1}}}}{\overline{{{a}_{3}}}\overline{{{d}_{2}}}}{{d}_{4}}|4\rangle$ & ${{e}_{4}}|0\rangle-\frac{\overline{{{a}_{1}}}{{a}_{4}}\overline{{{e}_{1}}}}{\overline{{{a}_{2}}}{{a}_{5}}
  \overline{{{e}_{2}}}}{{e}_{4}}|1\rangle-\frac{\overline{{{a}_{1}}}\overline{{{e}_{1}}}}{\overline{{{a}_{2}}}
  \overline{{{e}_{2}}}}{{e}_{4}}|3\rangle$\\
  \hline
  ${{f}_{7}}|1\rangle-\frac{{{f}_{3}}\overline{{{f}_{3}}{{f}_{4}}}}{\overline{{{f}_{1}}}{{f}_{2}}\overline{{{f}_{5}}}}
  {{f}_{7}}|2\rangle+\frac{{{f}_{3}}\overline{{{f}_{4}}}}{{{f}_{2}}\overline{{{f}_{5}}}}{{f}_{7}}|4\rangle$ & ${{b}_{4}}|0\rangle-\frac{{{a}_{3}}\overline{{{a}_{3}}}\overline{{{b}_{1}}}}{\overline{{{a}_{1}}}{{a}_{2}}
  \overline{{{b}_{2}}}}{{b}_{4}}|2\rangle+\frac{{{a}_{3}}\overline{{{b}_{1}}}}{{{a}_{2}}\overline{{{b}_{2}}}}{{b}_{4}}
  |4\rangle$ & ${{c}_{4}}|0\rangle-\frac{{{a}_{2}}\overline{{{a}_{2}}}\overline{{{a}_{5}}}
  \overline{{{c}_{1}}}}{{{a}_{3}}\overline{{{a}_{3}}}\overline{{{a}_{4}}}\overline{{{c}_{2}}}}{{c}_{4}}|1\rangle-
  \frac{\overline{{{a}_{2}}}\overline{{{c}_{1}}}}{\overline{{{a}_{3}}}\overline{{{c}_{2}}}}{{c}_{4}}|4\rangle$ & ${\color{red}|3\rangle}$ & ${{e}_{7}}|0\rangle+\frac{{{a}_{2}}{{a}_{4}}\overline{{{e}_{1}}}}{{{a}_{1}}{{a}_{5}}
  \overline{{{e}_{2}}}}{{e}_{7}}|1\rangle-\frac{\overline{{{e}_{1}}}}{\overline{{{e}_{2}}}}{{e}_{7}}|2\rangle$\\
  \hline
  ${{f}_{10}}|1\rangle+\frac{\overline{{{f}_{2}}}\overline{{{f}_{4}}}}
  {\overline{{{f}_{1}}}\overline{{{f}_{5}}}}{{f}_{10}}|2\rangle-\frac{\overline{{{f}_{4}}}}
  {\overline{{{f}_{5}}}}{{f}_{10}}|3\rangle$ & ${{b}_{7}}|0\rangle+\frac{\overline{{{a}_{2}}}\overline{{{b}_{1}}}}
  {\overline{{{a}_{1}}}\overline{{{b}_{2}}}}{{b}_{7}}|2\rangle-\frac{\overline{{{b}_{1}}}}{\overline{{{b}_{2}}}}
  {{b}_{7}}|3\rangle$ & ${{c}_{7}}|0\rangle+\frac{\overline{{{a}_{5}}}\overline{{{c}_{1}}}}{\overline{{{a}_{4}}}
  \overline{{{c}_{2}}}}{{c}_{7}}|1\rangle-\frac{\overline{{{c}_{1}}}}{\overline{{{c}_{2}}}}{{c}_{7}}|3\rangle$ & ${{d}_{7}}|0\rangle-\frac{{{a}_{3}}\overline{{{a}_{3}}}{{a}_{4}}\overline{{{d}_{1}}}}{{{a}_{1}}
  \overline{{{a}_{2}}}{{a}_{5}}\overline{{{d}_{2}}}}{{d}_{7}}|1\rangle-\frac{\overline{{{d}_{1}}}}{\overline{{{d}_{2}}}}
  {{d}_{7}}|2\rangle$ & ${\color{red}|4\rangle}$\\
  \hline
\end{tabular}}}
\vspace{0.3cm}

Among them~${{a}_{1}}, {{a}_{2}}, {{a}_{3}}, {{a}_{4}}, {{a}_{5}}, {{f}_{1}}, {{f}_{2}}, {{f}_{3}}, {{f}_{5}}, {{b}_{2}}, {{c}_{2}}, {{d}_{2}}, {{e}_{2}}$~ are not equal to 0.
\end{sidewaystable}

Furthermore, according to the orthogonality of the elements in each row, we have

\begin{equation}\label{A1}
\langle{{\Phi}_{10}}|{{\Phi}_{12}}\rangle=\overline{{{f}_{2}}}{{c}_{2}}+\overline{{{f}_{3}}}\frac{{{a}_{3}}}{{{a}_{2}}}{{c}_{2}}=0,
~~~~~~~~~~{{f}_{3}}=-\frac{\overline{{{a}_{2}}}}{\overline{{{a}_{3}}}}{{f}_{2}};
\end{equation}
\vspace{-0.4cm}
\begin{equation}\label{A2}
\langle{{\Phi}_{10}}|{{\Phi}_{13}}\rangle=\overline{{{f}_{1}}}{{d}_{2}}+\overline{{{f}_{3}}}\frac{{{a}_{3}}}{{{a}_{1}}}{{d}_{2}}=0,
~~~~~~~~~{{f}_{3}}=-\frac{\overline{{{a}_{1}}}}{\overline{{{a}_{3}}}}{{f}_{1}};
\end{equation}\vspace{-0.4cm}
\begin{equation}\label{A3}
\langle{{\Phi}_{10}}|{{\Phi}_{14}}\rangle=\overline{{{f}_{1}}}{{e}_{2}}+\overline{{{f}_{2}}}\frac{{{a}_{2}}}{{{a}_{1}}}{{e}_{2}}=0,
~~~~~~~~~{{f}_{2}}=-\frac{\overline{{{a}_{1}}}}{\overline{{{a}_{2}}}}{{f}_{1}}.
\end{equation}
Substituting Eq.~(\ref{A3}) into Eq.~(\ref{A1}) yields ~${{f}_{3}}=\frac{\overline{{{a}_{1}}}}{\overline{{{a}_{3}}}}{{f}_{1}}$,  this contradicts  with Eq.~(\ref{A2}).\vspace{0.2cm}

As a conclusion, there does not exist a non-classical idempotent QLS$(5)$.
\qed

\end{document}